\documentclass[pre,twocolumn,amsmath,amssymb]{revtex4}

\usepackage{graphicx}
\usepackage{dcolumn}
\usepackage{bm}

\begin{document}

\newcommand{\vspfigA}{\vspace{0.3cm}
   \rule{8.5cm}{0.3mm}
   \vspace{-0.2cm}
   }  
\newcommand{\vspfigB}{
   \rule{8.5cm}{0.3mm}
   \vspace{0.3cm}} 
\newcommand{\widthfigA}{0.35\textwidth} 
\newcommand{\widthfigB}{0.4\textwidth} 
\newcommand{\widthfigD}{0.85\textwidth} 
\newcommand{\widthfigE}{0.35\textwidth}

\newcommand{\bfGamma}{\mbox{\boldmath $\bf\Gamma$}} 



\title{
Localized behavior in the Lyapunov vectors 
for quasi-one-dimensional many-hard-disk  systems}

\author{Tooru Taniguchi and Gary P. Morriss}

\affiliation{School of Physics, University of New South Wales, Sydney, 
New South Wales 2052, Australia}

\date{\today}

\begin{abstract}

   We introduce a definition of a "localization width" whose logarithm is given by  
the entropy of the distribution of  particle component amplitudes in the 
Lyapunov vector. 
   Different types of localization widths are observed, for example, 
a minimum localization width where the components of only two 
particles are  dominant.
   We can distinguish a delocalization associated with a random 
distribution of particle contributions, a delocalization associated 
with a uniform distribution and a delocalization 
associated with a wave-like structure in the Lyapunov vector.
   Using the localization width we show that in quasi-one-dimensional 
systems of many hard disks there are two kinds of  dependence of
the localization width on the Lyapunov exponent index for the larger 
exponents: one is exponential, and the other is linear. 
   Differences, due to these kinds of localizations also 
appear in the shapes of the localized peaks of the Lyapunov vectors, 
the Lyapunov spectra and the angle between the spatial and 
momentum parts of the Lyapunov vectors. 
   We show that the Krylov relation for the largest Lyapunov 
exponent $\lambda\sim-\rho\ln\rho$ as a function of the density 
$\rho$ is satisfied (apart from a factor) in the same density region as the linear 
dependence of the localization widths is observed.  
   It is also shown that there are asymmetries in the 
spatial and momentum parts of the Lyapunov 
vectors, as well as  
in their $x$ and $y$-components. 

\vspace{0.2cm}
\noindent
Pacs numbers:
05.45.Jn, 
05.45.Pq, 
02.70.Ns, 
05.20.Jj 
\vspace{1cm}
\end{abstract}

\maketitle


\section{Introduction}

   The dynamical instability of a system is essentially connected  to 
the unpredictability of that system. 
   It is described by the time-evolution of the difference 
of two phase space vectors describing nearby trajectories, 
the so called the Lyapunov vector. 
   If the amplitude of the Lyapunov vector increases 
(decreases) rapidly in time, then the dynamics is unstable 
(stable) in the direction of the Lyapunov vector.  
   An unstable orbit implies that a part of the dynamics is 
unpredictable and a statistical treatment may be required. 
   The Lyapunov exponent, defined as the rate of exponential 
divergence or contraction of the amplitude of the Lyapunov 
vector with time, is the most frequently used indicator of 
the dynamical instability, and a system with a nonzero positive 
Lyapunov exponent is called chaotic. 
   The Lyapunov exponent is connected to the transport 
coefficients, such as the conductance and viscosity \cite{Gas98}. 
   In general, for each Lyapunov vector there is an individual 
Lyapunov exponent that is different in magnitude, and  
this leads to the concept of a sorted set of  
Lyapunov exponents, the so called  Lyapunov spectrum. 

   Although the significance of the \emph{amplitudes} of the 
Lyapunov vectors has been emphasized in the discussion of the dynamical 
instability, the Lyapunov vector itself including 
the information about its \emph{angle} can play an important 
role in chaotic dynamics \cite{Gas98,Gol87}. 
   For example, the Lyapunov vector was used to characterize 
a high-dimensional chaotic attractor \cite{Ike86} and 
the clustered motion in symplectic coupled map systems \cite{Kon92}.
   One may also mention that the wavelike structure of the Lyapunov 
vectors associated with the stepwise structure of the 
Lyapunov spectra in the small (in absolute value) Lyapunov 
exponent region has been reported in many-hard-disk systems 
\cite{Pos00,Eck00,Mcn01b,Tan02a}. 

   The localization of the Lyapunov vector for 
high-dimensional chaos, which we call the "Lyapunov localization" 
for convenience in this paper, is one of the chaotic behaviors 
which involves information on all the individual components of 
the Lyapunov vectors. 
   The Lyapunov localization appears as the spatial localization 
of the largest components of the Lyapunov vector and is generally
associated with the large Lyapunov exponents. 
   Such a phenomenon is reported in the 
Kuramoto-Sivashinsky model \cite{Man85}, 
in coupled map lattice systems \cite{Kan86,Gia91}, 
in a random matrix model \cite{Liv89}, 
in a high-dimensional symplectic map system \cite{Fal91} and 
in many-hard-disk systems \cite{Mil02,For02}. 
   However it was not so clear whether the Lyapunov 
localization has its origin in a randomness produced by 
the chaotic dynamics \cite{lep96} (like Anderson localization 
\cite{Lif88}) or it comes simply from a short range property of 
particle interactions.  
   Moreover physical meaning and significances of this 
phenomenon has not been well understood. 
   One of the few suggestions about the importance of the Lyapunov 
localization is that it may be related to the existence of  the
thermodynamic limit for the largest Lyapunov exponent \cite{Mil02}. 
   If a spatial localized region of the Lyapunov vector 
corresponding to the largest Lyapunov exponent is independent of 
the particle number $N$ in the thermodynamic limit, then 
the largest Lyapunov exponent can be $N$-independent. 
   This implies that the Kolmogorov-Sinai entropy, which is 
equal to the sum of all the positive Lyapunov exponents 
in closed systems, is an 
extensive quantity like the thermodynamic entropy.  
   Besides, this gives some supporting evidence to the existence 
of the thermodynamic limit of the Lyapunov spectrum, 
which has been a subject of study in many-particle chaotic systems 
\cite{Man85,Liv87,Sin96,Tan02b}. 

   The principal aim of this paper is to use the Lyapunov 
localization 
as an indicator of the chaotic properties of many-particle systems. 
   For quantitative considerations of the Lyapunov localization 
we use an entropy-like quantity (or information dimension) 
of a distribution function of the amplitudes of the Lyapunov 
vector components for each particle \cite{Fal91,Gia91}. 
   We introduce the "localization width" as the quantity 
whose logarithm is given by this entropy-like quantity. 
   The value of the localization width is in the range $[1,N]$, 
and indicates the number of particles contributing 
to the localized part of the Lyapunov vector components. 
   Using the localization width we can also distinguish different 
types of delocalized behaviors of the Lyapunov vectors such as 
a delocalization associated with a random 
distribution of particle component amplitudes of the Lyapunov vector, 
a delocalization associated with a uniform distribution 
and a delocalization associated with a wave-like structure 
in the Lyapunov vector. 
   As a concrete system to consider for the study of Lyapunov localization, 
we use a quasi-one-dimensional system consisting of many hard-disks, 
in which the system shape is so narrow as to exclude exchange 
of particle positions. 
   In this system the minimum value of the localization width is given 
by $2$, because particle interactions are given by collisions between 
\emph{two} particles. 
   In the quasi-one-dimensional systems each particle can interact 
only with its nearest neighbor particles, so the numerical calculation 
is less time-consuming than for a fully two-dimensional hard-disk system 
in which each particle can collide with any other particle. 
   The quasi-one-dimensional system also has an advantage in 
that the roles of the coordinate directions are strongly separated. 
  In our system the narrow direction (the $y$-direction) of the 
rectangle is very much different from the longer orthogonal direction 
(the $x$-direction). 
   Another advantage of the use of the quasi-one-dimensional system 
is that there is a wider region of stepwise structure in the Lyapunov spectrum 
compared to a square two-dimensional system \cite{Tan02c}. 
   This is a noticeable point because in this paper we 
show that the localization width is 
an indicator not only of the Lyapunov localization but also 
of the stepwise structure of the Lyapunov spectrum. 
   Calculating the localization width in quasi-one-dimensional 
systems we show that there are two kinds of dependence of 
the Lyapunov index on the localization widths:  
an exponential dependence [$\mathcal{ED}$] 
and a linear dependence [$\mathcal{LD}$]. 
   (Here we define the Lyapunov spectrum as the 
set $\{\lambda^{(1)},\lambda^{(2)},\cdots,\lambda^{(4N)}\}$ 
of the Lyapunov exponents satisfying the condition 
$\lambda^{(1)}\geq\lambda^{(2)}\geq\cdots\geq\lambda^{(4N)}$ 
and introduce the the Lyapunov index as the ordering number 
of the exponents in the Lyapunov spectrum.) 
   The exponential dependence [$\mathcal{ED}$] of the localization 
width appears at any particle density, and shows a tail 
in the spatial localized shape of the Lyapunov vector. 
   On the other hand the linear dependence [$\mathcal{LD}$] 
of the localization widths as a function of Lyapunov index appears 
only in cases of low particle density, 
and is characterized by a sharp rectangular 
localized shape of the Lyapunov vector in space. 
   Next, we consider the effects of the these two kinds 
of Lyapunov localizations on other quantities, 
such as the shape of the Lyapunov spectra, the angle between the 
spatial part and the momentum part of the Lyapunov vector, 
and the amplitudes of the spatial part of the Lyapunov vector, etc. 
   In particular, it is shown that apart from a prefactor 
the Krylov relation for the largest Lyapunov exponent 
$\lambda\sim-\rho\ln\rho$ 
as a function of the particle density 
$\rho$ is satisfied in the same density region 
in which the linear dependence [$\mathcal{LD}$] 
of the localization widths appears. 
   We also demonstrate a relation between 
localized regions of the Lyapunov vectors and the positions of 
colliding particles, suggesting that the Lyapunov localization 
comes from the short range interaction of the particles. 
   We compare some of our results in quasi-one-dimensional 
systems with the square system with the same area, and show 
that the two kinds of localizations also appear in 
the square case as well. 

   As the second aim of this paper we investigate how 
the particle density and system shape affect the 
Lyapunov vector components. 
   We show that the spatial part 
and the momentum part of the Lyapunov vectors are 
in almost the same direction at high density, 
whereas they are rather close to orthogonal 
in low density cases, 
especially in the region of 
the exponential dependence [$\mathcal{ED}$] of the 
localization widths as a function of Lyapunov index. 
   The amplitudes of the spatial part of the 
Lyapunov vectors are larger (smaller) than the corresponding 
momentum part in low density cases (high density cases).
   We also demonstrate that gaps appear in the amplitudes 
of the $x$-components and the $y$-components of the 
spatial part and the momentum part 
of the Lyapunov vectors, 
because of the difference of the roles of the directions in the 
quasi-one-dimensional system.  
   In these amplitudes of the Lyapunov vector components 
we can also see effects of the stepwise structure of the 
Lyapunov spectra. 

   The outline of this paper is as follows.
   In Sect. \ref{LocalizationWidth} we compare some quantities 
that characterize the Lyapunov localization, and discuss the
relative merits of the localization width 
comparing it with the other quantities. 
   The relation between the localized region of Lyapunov vectors 
and the position of colliding particles is demonstrated. 
   In Sect. \ref{Locza-Density} we show that there are 
two kinds of Lyapunov localizations.  
   These two Lyapunov localizations are 
distinguished by their Lyapunov index dependences and 
the shapes of the localized peaks of the 
Lyapunov vectors. 
   In Sect. \ref{spectraandangles} we investigate 
the effects of the two kinds of Lyapunov localizations on the 
shape of the Lyapunov spectra, and angles and amplitudes 
of the Lyapunov vectors components, etc. 
   The Lyapunov index dependence of the $x$ and $y$-components 
of the spatial part and the momentum part of the Lyapunov vectors are 
shown. 
   In Sect. \ref{densitydependenceLmax} we investigate 
the density dependences 
of quantities associated with the largest Lyapunov exponent 
such as the largest Lyapunov exponent itself, the angle of 
the Lyapunov vectors and the localization width, etc., 
and specify the density region 
in which the linear dependence [$\mathcal{LD}$] 
of the localization width on Lyapunov index appears. 
   The region of Lyapunov indices, in which the linear dependence 
[$\mathcal{LD}$] of the Lyapunov 
widths appears, is connected to the region in which  
the Krylov relation for the density dependence 
of the largest Lyapunov exponent is satisfied. 
   Sect. \ref{ConclusionRemarks} is our conclusions and further
remarks . 
   In Appendix \ref{appen0} we give a derivation of 
an inequality for the localization width. 
   In Appendix \ref{appenA} we compare 
some results for the quasi-one-dimensional systems in the text 
with those of corresponding square systems.


\section{Localization Width of Lyapunov Vectors}
\label{LocalizationWidth}

   We consider the quasi-one-dimensional system which consists 
of $N$ hard-disks with radius $R$. 
   The shape of the system is a two-dimensional rectangle 
satisfying the condition 

\begin{eqnarray}
   RN \sqrt{3} < L_{x} \;\; \mbox{and} \;\; 
   2R < L_{y} < 4R, 
\label{qua1dimcon}\end{eqnarray}

\noindent with width $L_{x}$ and height $L_{y}$ .
   Condition (\ref{qua1dimcon}) means that the system is 
so narrow that the positions of particles cannot be exchanged. 
   The systems satisfying this condition are referred to as 
quasi-one-dimensional systems in this paper. 
   The schematic illustration of such a system is given in 
Fig. \ref{qua1dim}. 
   The quasi-one-dimensional system was used 
to discuss the stepwise 
structure of the Lyapunov spectra and the associated 
wave-like structure of the Lyapunov vectors in Ref. \cite{Tan02c}. 

\begin{figure}[!htb]
\vspfigA
\caption{Quasi-one-dimensional system.  
   The system shape is so narrow and the exchange of particle 
   positions are excluded.}
\vspfigB
\label{qua1dim}\end{figure}  

   In this paper we consider the case where the parameter values are 
given by; the radius of particles is $R=1$, the mass of particles is $M=1$, 
and the total energy of system is $E=N$. 
   The system size is given by 
$L_{y} = 2R(1+10^{-6})$ and  $L_{x} = NL_{y}(1+d)$ 
satisfying the condition (\ref{qua1dimcon}) 
with a constant $d$, (except in Appendix \ref{appenA} 
where we consider square systems.) 
   Roughly speaking the factor $1+d$ is the averaged ratio of the 
distances that each particle can move in the $x$-direction 
and $y$-direction. 
   The constant $d$ is connected to the density 
   
\begin{eqnarray}
   \rho\equiv \frac{N \pi R^{2}}{L_{x}L_{y}} 
   = \frac{\pi R^{2}}{(1+d)L_{y}^{2}}
\label{densi}\end{eqnarray}

\noindent as $d=\pi R^{2}/(\rho L_{y}^{2}) - 1$ \cite{noteDensi}.

   The quasi-one-dimensional many-hard-disk system has many 
advantages for numerical investigations of the dynamical 
properties of many-particle systems.
   First, in many-hard-disk systems the free flight part of 
the dynamics is integrable, so the actual numerical calculations 
of phase space dynamics and tangent space dynamics are simply 
described by multiplications of the time-evolutional matrices 
corresponding to free flight and collision. 
   Second, in the quasi-one-dimensional system defined by 
Eq. (\ref{qua1dimcon}) each particle can only collide with its two 
nearest-neighbor particles, so we do not need to search 
every particle pair to find its colliding particles 
in numerical calculations.
   These points lead to a faster numerical 
calculation of the dynamical properties (especially Lyapunov spectra 
and Lyapunov vectors) than for other many-particle systems,  
such as fully two-dimensional particle systems (in which 
any pair of particles can collide), or with particles 
with soft core interactions.
   As another advantage of the quasi-one-dimensional system is 
that the roles of the $x$ and $y$-directions are separated, 
so we can investigate how such a separated role for the directions 
effects for example,  the Lyapunov exponents and Lyapunov 
vectors. 
   In the quasi-one-dimensional system 
the information of the particle positions can be roughly 
replaced by the particle indices, 
so we can discuss the dynamics of quantities 
by using three-dimensional 
graphs as functions of collision number and particle index, 
which is much simpler than the fully two-dimensional system 
that requires four-dimensional graphs as functions of the time, 
the $x$ and $y$-components of particle positions. 
   Finally in quasi-one-dimensional many-hard-disks, 
we get a wider stepwise region of the Lyapunov spectra 
rather than in the fully two-dimensional square systems.  
   This was an essential point in the study of
 the stepwise structure of Lyapunov spectra 
and the associated wave-like structure of the Lyapunov vectors 
in systems with small numbers of particle in Ref. \cite{Tan02c}. 

   We introduce the Lyapunov vector as 
$\delta\bfGamma^{(n)}(t) 
\equiv (\delta\bfGamma_{1}^{(n)}(t),
\delta\bfGamma_{2}^{(n)}(t),\cdots,
\delta\bfGamma_{N}^{(n)}(t))$ 
corresponding to the $n$-th Lyapunov exponents $\lambda^{(n)}$ 
at time $t$. 
   Here, $\delta\bfGamma_{j}^{(n)}(t)$ is the Lyapunov 
vector component corresponding to the $j$-th particle and the 
Lyapunov exponent $\lambda^{(n)}$ at time $t$. 
   We define the normalized amplitude $\gamma_{j}^{(n)}(t)$ of 
the Lyapunov vector components $\delta\bfGamma_{j}^{(n)}(t)$ 
by 

\begin{eqnarray}
   \gamma_{j}^{(n)}(t) \equiv 
   \frac{\left| \delta\bfGamma_{j}^{(n)}(t) \right|^{2}}
   {\sum_{k=1}^{N} 
   \left| \delta\bfGamma_{k}^{(n)}(t) \right|^{2}} ,
\label{lyaamp}\end{eqnarray}

\noindent so that the conditions 

\begin{eqnarray}
   \sum_{j=1}^{N} \gamma_{j}^{(n)}(t) = 1, 
   \label{lyaampB}
   \\
   0\leq \gamma_{j}^{(n)}(t) \leq 1
\label{lyaampC}\end{eqnarray}

\noindent are automatically satisfied. 
   Fig. \ref{loczat}(a) and (b) are two typical behaviors 
of the quantity $\gamma_{j}^{(1)}(t)$ corresponding to largest 
Lyapunov exponent as functions of the collision number 
$n_{t}$ and the particle index $j$ in the case of (a) $d=1.5$ 
(or density $\rho=0.314\cdots$) and 
(b) $d=10^{5}$ (or density $\rho=0.00000785\cdots$). 
   Here we take the particle index $j$ so that the $x$-component 
$q_{xj}(t)$ of the $j$-th particle satisfies the condition 
$q_{x1}(t) < q_{x2}(t)< \cdots <q_{xN}(t)$, 
and the data is taken every two collisions.
   (It should be noted that the collision number 
$n_{t}$ is related to the time $t$ approximately 
by multiplying by the mean free time, 
and in the quasi-one-dimensional system the particle index 
$j$ has a similar meaning to the $x$-component 
of the particle position.)
   These graphs show clearly that a non-zero value of the quantity 
$\gamma_{j}^{(1)}(t)$ is localized in a small spatial 
region involving a few particles. 
   Localized peak positions stay in almost the same position 
over several tens of collisions, then seems to hop to another position. 
   Detail of the localized behaviors of the quantity 
$\gamma_{j}^{(1)}(t)$,
especially the differences between Figs. \ref{loczat}(a) and (b), 
will be discussed in the following sections and are the main purpose 
of this paper. 

\begin{figure}[!htb]
\vspfigA
\caption{
      Localized behaviors of the amplitudes 
   $\gamma_{j}^{(1)}$ 
   of the normalized Lyapunov vector components of 
   the $j$-th particle corresponding to the largest 
   Lyapunov exponent $\lambda^{(1)}$ as functions of 
   the collision number 
   $n_{t}$ and the particle index $j$ 
   in a quasi-one-dimensional system of $N=50$. 
   (a) The high density case of $d=1.5$. 
   (b) The low density case of $d=10^{5}$.}
\vspfigB
\label{loczat}\end{figure}  

   To obtain our numerical results we use the algorithm 
developed by Benettin et al. \cite{Ben76} and 
Shimada et al. \cite{Shi79} and so on.  
   This algorithm is characterized by intermittent rescaling and 
renormalization of Lyapunov vectors, which can be done 
after each particle collision in a many-hard-disk system. 
   Other articles such as Refs. \cite{Ben80a,Ben80b,Del96} should be 
referred to for more details of this algorithm 
and the Lyapunov vector dynamics of many-hard-disk systems. 

   Now we discuss methods to characterize the strength of the 
localization of the Lyapunov vectors 
like those in Fig. \ref{loczat}(a) and (b) 
in a quantitative sense. 
   By analogy to the localization length 
used in Anderson localization \cite{Lif88} 
it may be suggested that the strength 
of the Lyapunov localization can be 
characterized by a localization length 
$\Omega^{(n)}$ defined by 

\begin{eqnarray}
   \left[\Omega^{(n)}\right]^{-1} \equiv \lim_{j\rightarrow\infty} 
   \lim_{N\rightarrow\infty} 
   j^{-1} \left\langle\ln\gamma_{j}^{(n)}(t)\right\rangle
\label{LoczatLengt}\end{eqnarray}

\noindent as the number of particles $N$ goes to infinity, namely 
in the thermodynamic limit. 
   Here we use the bracket $\langle{\cal X}\rangle$ to signify the 
long time average of the time-dependent quantity ${\cal X}$. 
   (Note that in the definition (\ref{LoczatLengt}) of 
the localization length $\Omega^{(n)}$ we used the fact that 
the system is quasi-one-dimensional. 
   In more general cases like a fully two-dimensional system 
the limit $j\rightarrow\infty$ in Eq. (\ref{LoczatLengt}) 
must be replaced by the limit 
of the amplitude of the spatial coordinate.)
   This quantity $\Omega^{(n)}$ characterizes the Lyapunov 
localization as its spatial exponential decay rate. 
   However this quantity may not be suitable to characterize the 
localization like that in Fig. \ref{loczat}(b) which does not have an 
exponential decay. 
   Besides, this quantity requires the thermodynamic limit 
$N\rightarrow\infty$. 
The numerical calculation of the 
Lyapunov spectrum and Lyapunov vectors 
for many-particle systems is very time-consuming 
and has so far only been reported for 
systems  of about 1000 particles  \cite{For02}, 
so it may be rather difficult to estimate the 
quantity $\Omega^{(n)}$ defined by Eq. (\ref{LoczatLengt}) numerically. 

   Another method to characterize the Lyapunov localization 
was proposed in Ref. \cite{Mil02}. 
   In this method we first introduce a parameter 
threshold $\Theta\in(0,1)$ and 
define the quantity $C^{(n)}_{\Theta}$ as 

\begin{eqnarray}
   C^{(n)}_{\Theta} \equiv \min_{x} 
   \left\{x \; ; \;\; \Theta <  \sum_{j=1}^{N x} 
      \left\langle 
      \tilde{\gamma}_{j}^{(n)}(t)
   \right\rangle
   \right\}
\label{LoczatC}\end{eqnarray} 

\noindent with an integer $N x$ and the sorted set $\{ 
\tilde{\gamma}_{1}^{(n)}(t),\tilde{\gamma}_{2}^{(n)}(t),
$ $\cdots,\tilde{\gamma}_{N}^{(n)}(t)\} 
= \{ 
\gamma_{1}^{(n)}(t),\gamma_{2}^{(n)}(t),
$ $\cdots,\gamma_{N}^{(n)}(t)\}$ 
satisfying the condition 
$\tilde{\gamma}_{1}^{(n)}(t)\geq\tilde{\gamma}_{2}^{(n)}(t)\geq
\cdots\geq\tilde{\gamma}_{N}^{(n)}(t)$. 
   This quantity $C^{(n)}_{\Theta}$ is a measure of the 
number of particle components actively contributing to 
a localized part of the Lyapunov vector for 
the $n$-th Lyapunov exponent $\lambda^{(n)}$. 
   This quantity $C^{(n)}_{\Theta}$ does not require the 
thermodynamic limit, and can be suitable for both types of the 
Lyapunov localizations shown in Fig. \ref{loczat}. 
   However this quantity is a function of an artificial threshold 
$\Theta$, which cannot be determined by the dynamics itself. 

   In this paper we discuss the Lyapunov localization by  
using an entropy-like quantity for the amplitude distribution  
$\gamma_{j}^{(n)}(t)$ of the normalized Lyapunov vector elements, 
namely the entropy-like quantity $S^{(n)}$ defined by

\begin{eqnarray}
   S^{(n)} \equiv - \sum_{j=1}^{N} \left\langle
   \gamma_{j}^{(n)}(t) \ln \gamma_{j}^{(n)}(t) \right\rangle,  
\label{LoczatEntro}\end{eqnarray} 

\noindent noting that the quantity 
$\gamma_{j}^{(n)}(t)$ satisfies the conditions 
(\ref{lyaampB}) and (\ref{lyaampC}) so it can be 
regarded as a kind of probability distribution function. 
   Using this quantity $S^{(n)}$ 
we introduce the quantity $\mathcal{W}^{(n)}$ as 

\begin{eqnarray}
   \mathcal{W}^{(n)} \equiv \exp\{S^{(n)}\}, 
\label{LoczatWidth}\end{eqnarray} 

\noindent which we call the $n$-th 
"localization width" corresponding to 
the $n$-th Lyapunov exponent $\lambda^{(n)}$. 
   This localization width has already been used to discuss the Lyapunov 
localization in a coupled map lattice model \cite{Gia91} 
and a high-dimensional symplectic map system \cite{Fal91}. 
   
   To understand the physical meaning of the localization width 
$\mathcal{W}^{(n)}$ defined by 
Eq. (\ref{LoczatWidth}) it is useful to discuss 
some of its properties. 
   The first property of the localization width 
is the inequality
   
\begin{eqnarray}
   1 \leq \mathcal{W}^{(n)} \leq N . 
\label{LoczatWidthPropa}\end{eqnarray} 

\noindent The derivation of the inequality 
(\ref{LoczatWidthPropa}) 
is given in Appendix \ref{appen0}. 
   The second property of the localization width 
$\mathcal{W}^{(n)}$ is that 
the equality $\mathcal{W}^{(n)} = 1$ is satisfied 
only when one of the quantities $\gamma_{j}^{(n)}(t)$, 
$j=1,2,\cdots N$ 
is equal to 1, namely in the most localized case, and the equality 
$\mathcal{W}^{(n)} = N$ is satisfied 
only when all of these quantities 
are equal to each other, namely 
$\gamma_{1}^{(n)}(t)=\gamma_{2}^{(n)}(t) 
=\cdots=\gamma_{N}^{(n)}(t) = 1/N$, 
namely in the most delocalized case 
(See Appendix \ref{appen0} about this point.). 
   These properties suggest that the localization width 
$\mathcal{W}^{(n)}$ also indicates the number of particles 
contributing to the localized part of Lyapunov vector, which is analogous 
to the quantity $N C^{(n)}_{\Theta}$ derived 
from Eq. (\ref{LoczatC}). 
   The third property of the localization width comes from 
the symplectic structure of the Hamiltonian dynamics. 
   To discuss this property we note the conjugate relation 
 \cite{Gas98}

\begin{eqnarray}
   \delta\bfGamma_{j}^{(4N-n+1)} 
   &\equiv&
   \left(\begin{array}{c}
      \delta \mathbf{q}_{j}^{(4N-n+1)} \\
      \delta \mathbf{p}_{j}^{(4N-n+1)}
   \end{array}\right)  \nonumber \\
   &=&
   \xi^{(n)} \left(\begin{array}{c}
      \delta \mathbf{p}_{j}^{(n)} \\
      - \delta \mathbf{q}_{j}^{(n)}
   \end{array}\right)  
\label{Sympl}\end{eqnarray} 

\noindent for the Lyapunov vector component 
$\delta\bfGamma_{j}^{(4N-n+1)}$ of the $j$-th particle 
corresponding to the $(4N-n+1)$-th Lyapunov exponent 

\begin{eqnarray}
   \lambda^{(4N-n+1)}= - \lambda^{(n)},  
\label{Sympl3}\end{eqnarray} 

\noindent $n=1,2,\cdots,2N$, 
where $\delta\mathbf{q}_{j}^{(n)}$ and 
$\delta\mathbf{p}_{j}^{(n)}$
are the spatial part and the momentum part of 
the Lyapunov vector $\delta\bfGamma_{j}^{(n)}$ of 
the $j$-th particle, respectively, and $\xi^{(n)}$ is a constant 
depending on the Lyapunov index only. 
   (As a remark, if we use the Benettin algorithm to calculate 
the Lyapunov spectra, which is characterized by the intermittent 
normalization (as well as the rescaling) of Lyapunov vectors, 
the factor $\xi^{(n)}$ in Eq. (\ref{Sympl}) is given 
by $+1$ or $-1$ depending on the Lyapunov index $n$.)
   From Eqs. (\ref{lyaamp}) and (\ref{Sympl}) we derive 

\begin{eqnarray}
   \gamma_{j}^{(4N-n+1)}(t) = \gamma_{j}^{(n)}(t)
\label{Sympl2}\end{eqnarray} 

\noindent at any time $t$, where $j$ is the particle index 
and $n = 1,2,\cdots, 2N$ is the Lyapunov index.  
   Eqs. (\ref{LoczatEntro}), (\ref{LoczatWidth}) 
and (\ref{Sympl2}) lead to the third property of the 
localization width 
   
\begin{eqnarray}
   \mathcal{W}^{(4N-n+1)} = \mathcal{W}^{(n)}. 
\label{LoczatWidthSympl}\end{eqnarray} 

\noindent This is the conjugate property for the localization 
width $\mathcal{W}^{(n)}$.
   
   Some of the advantages and disadvantages of the use of the 
localization width 
$\mathcal{W}^{(n)}$ to discuss the Lyapunov localizations 
are as follows. 
   An advantage of the localization width $\mathcal{W}^{(n)}$ 
is that its calculation  
does not require the thermodynamic limit, different from  
the localization length $\Omega^{(n)}$ defined by 
Eq. (\ref{LoczatLengt}). 
   Beside, the localization width $\mathcal{W}^{(n)}$ 
is applicable to the 
Lyapunov localization seen in Fig. \ref{loczat}(b), whereas 
the localization length $\Omega^{(n)}$ requires that the tail 
of the quantity $\gamma_{j}^{(n)}(t)$ decay exponentially in space. 
   Moreover the localization width $\mathcal{W}^{(n)}$ does not require 
an artificial parameter like the threshold $\Theta\in(0,1)$ in 
the quantity $C^{(n)}_{\Theta}$ defined by Eq. (\ref{LoczatC}), 
so it can be determined from the dynamics only. 
   On the other hand, a disadvantage of the localization width 
$\mathcal{W}^{(n)}$ is that using this quantity, for example, 
we cannot distinguish between \emph{one} peak of height $\alpha$ and 
width $2\beta$ with constants $\alpha$ and $\beta$ and 
\emph{two} peaks of height $\alpha$ and width 
$\beta$ which should be recognized as different localized behaviors.   
   Therefore, in principle we have to look at the concrete shapes 
of localized peaks to check the relation between this localization width 
and the peak width in actual numerical calculations, 
even if we calculate the localization width of the Lyapunov vector. 
   The quantity $C^{(n)}_{\Theta}$ defined by Eq. (\ref{LoczatC}) 
also has the same disadvantage.
 
\begin{figure}[!htb]
\vspfigA
\caption{
      Normalized localization widths $\mathcal{W}^{(n)}/N$ 
   in the case of $d=0.5$. 
      The circles, triangles and squares 
   correspond to the cases of $N=25$, $50$ and $100$, 
   respectively. 
      The broken line interrupted by dots, 
   the solid line and the broken lines (and the dotted line) 
   correspond to  
   $\mathcal{W}^{(n)} = \mathcal{W}_{wav} (\approx 0.736N)$, 
   $\mathcal{W}_{ran} (\approx 0.651N)$ 
   and $\mathcal{W}_{min} (=2)$, 
   respectively. 
      (a) Full scale of the normalized localization widths as 
   functions of the normalized Lyapunov index $n/(2N)$. 
      (b) An enlarged graph of the normalized localization widths 
   corresponding to the small positive 
   Lyapunov exponent region as functions of 
   the Lyapunov index $n$. 
      Symbols with a grey-filled background correspond 
   to the two-point steps of the Lyapunov spectra 
   which are indicated by similar circles in Fig. \ref{lypuaNdepen}.
      Symbols surrounded by a rectangle of broken lines 
   correspond to the four-point steps of the Lyapunov spectra 
   which are indicated by similar rectangles surrounding symbols 
   in Fig. \ref{lypuaNdepen}.
   }
\vspfigB
\label{loczwidndep}\end{figure}  

   Fig. \ref{loczwidndep} are examples of graphs of 
localization widths as functions of Lyapunov index
in quasi-one-dimensional systems. 
   Here we have plotted the localization widths 
$\mathcal{W}^{(n)}/N$ normalized by the 
particle number $N$ for quasi-one-dimensional systems of  
size $N=25$ (circles), $N=50$ (triangles) and $N=100$ 
(squares) where $d=0.5$ 
(density $\rho=0.523\cdots$). 
   In order to take the time-average in Eq. (\ref{LoczatEntro}) 
to calculate the localization width, 
we sample the quantities 
$\gamma_{j}^{(n)}(t) \ln\gamma_{j}^{(n)}(t)$ 
just after collisions over $1000N$ collisions and take 
arithmetic averages of them.
(In this paper we always calculate localization widths 
in this way.)
   Fig. \ref{loczwidndep}(a) is the full scale graph 
of the localization widths as functions of the normalized 
Lyapunov index $n/(2N)$, and  Fig. \ref{loczwidndep}(b) 
is a part of the localization widths corresponding 
to the small positive 
Lyapunov exponent region as functions of the 
Lyapunov index $n$. 
   (Note that we use the normalized Lyapunov index $n/(2N)$ for 
the $x$-axis in Fig. \ref{loczwidndep}(a), 
which is different from the $x$-axis as the Lyapunov index $n$ 
itself in Fig. \ref{loczwidndep}(b).) 
   To plot the values of the localization widths 
$\mathcal{W}^{(n)}$, 
$j=2N+1,2N+2,\cdots,4N$ is omitted because of the conjugate 
relation (\ref{LoczatWidthSympl}).  
   In the region of large positive Lyapunov exponents 
the localization width $\mathcal{W}^{(n)}$ is an monotonically 
increasing function of the Lyapunov index $n$. 
   This implies that localized behavior of the 
Lyapunov vector is stronger in the large (in absolute value) 
Lyapunov exponent region.  
   The shape of the localization width in this region 
is similar qualitatively to the shape given by the quantity 
$C^{(n)}_{\Theta}$ defined by Eq. (\ref{LoczatC}) 
which was calculated in a square system \cite{For02}. 
   Figure \ref{loczwidndep}(a) also shows that the 
value of the normalized Lyapunov 
localization widths $\mathcal{W}^{(n)}/N$ decrease as 
a function of particle number $N$ in the region where 
Lyapunov spectra change smoothly. 
   Noting that the localization width has a lower bound 
($\mathcal{W}^{(n)}/N \geq 1/N$), this 
suggests the existence of the thermodynamic limit for the 
spectrum of the localization widths.

\begin{figure}[!htb]
\vspfigA
\caption{Lyapunov spectra normalized by 
   the largest Lyapunov exponent $\lambda^{(1)}$ as functions of 
   the Lyapunov index $n$ for the case of $d=0.5$. 
      The circles, triangles and squares 
   correspond to the cases of $N=25,50$ and $100$, 
   respectively. 
      Symbols with a grey-filled background 
   correspond to symbols with similar circles 
   in the graph \ref{loczwidndep}(b).
      Symbols surrounded by a rectangle of broken lines 
   correspond to symbols surrounded by similar rectangles
   in the graph \ref{loczwidndep}(b).
      Inset: The full scale graph of the positive branch of 
   the Lyapunov spectra as functions of 
   the normalized Lyapunov index $n/(2N)$.}
\vspfigB
\label{lypuaNdepen}\end{figure}  

   Another important point in Fig. \ref{loczwidndep} 
is the connection with the stepwise structure 
of Lyapunov spectra. 
   Fig. \ref{lypuaNdepen} is the Lyapunov spectra normalized by 
the largest Lyapunov exponent $\lambda^{(1)}$ as a function of 
the Lyapunov index $n$ in the case of 
$N=25$ (circles), $N=50$ (triangles) and $N=100$ 
(squares) with $d=0.5$. 
   These three cases correspond to the three cases 
of the localization widths in Fig. \ref{loczwidndep}. 
   In this figure the stepwise structures of Lyapunov spectra 
appear in the small Lyapunov exponent region. 
   The values of the largest Lyapunov exponents are 
$\lambda^{(1)} \approx 1.28$ for $N=25$, 
$\lambda^{(1)} \approx 1.31$ for $N=50$, 
$\lambda^{(1)} \approx 1.31$ for $N=100$. 
   The inset of Fig. \ref{lypuaNdepen} is the full scale graphs 
of the positive branch of the Lyapunov spectra 
normalized by the largest Lyapunov exponents as functions 
of the normalized Lyapunov index $n/(2N)$ 
(Note again that we use the different horizontal axes 
in the main graphs and the inset in Fig. \ref{lypuaNdepen}.), 
and show that the global shapes of these graphs are similar and 
independent of the particle number $N$.
   Comparing Fig. \ref{lypuaNdepen} with Fig. \ref{loczwidndep}(b) 
for the localization widths in the same region of 
the Lyapunov index $n$, we notice that 
the localization widths $\mathcal{W}^{(n)}/N$ 
normalized by the particle number $N$ 
corresponding to the clear two-point steps of the Lyapunov 
spectra take almost same value which is almost independent of 
the sequence of the steps and the particle number $N$  
   (See the symbols with a grey-filled background 
in Figs. \ref{loczwidndep}(b) and \ref{lypuaNdepen} 
about this point.).  
   A similar thing can be seen for the four-point steps 
of the Lyapunov spectra. 
   (See the symbols surrounded by a rectangle of broken lines 
in Figs. \ref{loczwidndep}(b) and \ref{lypuaNdepen}.)
   One may also notice that 
the region of the localization widths 
corresponding to the stepwise region of the Lyapunov spectra, 
which is around the region $n/(2N) > 0.8$ 
of the Lyapunov index $n$ in Figs. \ref{loczwidndep} 
approximately, 
can be distinguished from the other region. 
    These points about the localization widths 
that connect with the stepwise structure of the 
Lyapunov spectra are other merits of the use of 
the localization width to discuss the behavior of the Lyapunov vectors.

\begin{figure}[!htb]
\vspfigA
\caption{
      Localized regions (grey-filled regions 
   surrounded by solid lines) 
   of Lyapunov vectors $\delta\bfGamma^{(1)}$
   and the numbers of colliding particle pairs 
   (as pairs of black-filled circles) as functions of 
   the collision number $n_{t}$ and the particle 
   number $j$
   in a quasi-one-dimensional system of $N=50$ and $d=10^{5}$. 
      The solid line is the contour plot $\gamma_{j}^{(1)}=0.3$ 
   of the amplitudes of the normalized Lyapunov vector components 
   $\gamma_{j}^{(1)}$ of the $j$-th particle corresponding 
   to the largest Lyapunov exponent $\lambda^{(1)}$, 
   which corresponds to the three dimensional plot 
   given in Fig. \ref{loczat}(b).  
   }
\vspfigB
\label{loczaamplicolli}\end{figure}  

   Before finishing this section 
it is valuable to summarize some specific values 
of the localization width $\mathcal{W}^{(n)}$ 
which have clear physical meanings. 
   The first value of the localization width 
is $\mathcal{W}^{(n)} = \mathcal{W}_{max} 
\equiv N$ which means that the amplitudes 
$|\delta\bfGamma_{j}^{(n)}|$ 
of the Lyapunov vector components for each particle take the same values 
as each other, as mentioned before. 
   The Lyapunov localization close to this value actually occurs 
in one of the zero-Lyapunov exponent, 
as shown in $\mathcal{W}^{(2N)}/N$ in Fig. \ref{loczwidndep}(a). 
   The second value is the lower bound of the Lyapunov localization: 
$\mathcal{W}^{(n)}\geq\mathcal{W}_{min}\equiv 2$.  
   This is a little stronger condition than the inequality 
$\mathcal{W}^{(n)}\geq 1$ which we have already shown 
in the inequality (\ref{LoczatWidthPropa}). 
   This value of the localization width is shown in Fig. \ref{loczwidndep}
as the dotted line 
(the three lines of $\mathcal{W}^{(n)}/N = 2/N 
= 0.08$ for $N=25$, $2/N = 0.04$ for $N=50$, and 
$2/N = 0.02$ for $N=100$). 
   This lower bound for the localization width comes from  
the fact that particle collisions occur between \emph{two} 
particles, and is partly supported by the fact that 
the width of the 
the amplitudes  $\gamma_{j}^{(n)}$ of the normalized 
Lyapunov vector components of the $j$-th particle corresponding 
to the $n$-th Lyapunov exponent $\lambda^{(n)}$ is almost 
$2$ in the low density limit and in 
large Lyapunov exponents, for example, 
as shown in Fig. \ref{loczat}(b).   
   It should also be emphasized that there is a relation 
between the localized region of a Lyapunov vector 
and positions of colliding particle pairs. 
   To show this point, in Fig. \ref{loczaamplicolli} 
we plot colliding particle pairs 
as well as a contour plot of the amplitude $\gamma_{j}^{(1)}$ 
as functions of the collision number $n_{t}$ and the particle 
number $j$ in a quasi-one-dimensional system of $N=50$ and 
$d=10^{5}$, which correspond to the three dimensional plot of 
the Lyapunov localization given in Fig. \ref{loczat}(b).  
   It is clear from this figure that particle collisions occur 
at the beginning of the sharp rectangular shapes of the localization 
of the amplitudes $\gamma_{j}^{(n)}$. 
   This suggests that the Lyapunov localization 
comes from a short interaction range of particle-particle 
interactions. 
   The minimum value of the localization width 
will also be discussed in the subsections 
\ref{loczatinLowDens} and \ref{localizationwidthcorresponding} 
in this paper. 
   The third value of the localization width is the value for  
the case in which the amplitudes $\gamma_{j}^{(n)}(t)$, 
$k=1,2,\cdots,N$ are distributed randomly with an equal probability 
(the solid line in Fig. \ref{loczwidndep}). 
   It is given by $\mathcal{W}^{(n)} = \mathcal{W}_{ran} 
\approx0.651N$ approximately, and gives an 
upper bound on the localization widths of the Lyapunov 
index corresponding to the continuos part of the Lyapunov spectrum 
 in many cases \cite{noteWran}. 
   The fourth value of the localization width is the value obtained for
 a wavelike structure in the Lyapunov vector. 
   The broken line in Fig. \ref{loczwidndep} is given as  
the localization width of $\gamma_{j}^{(n)}(t) = 
\langle\langle|\alpha_{j}^{(n)}\sin(2\pi jn/N+\beta)|^{2} 
\rangle\rangle$ where the constant $\alpha_{j}^{(n)}$ is 
determined by the normalization condition of 
$\gamma_{j}^{(n)}(t)$ and $\langle\langle\cdots\rangle\rangle$
means the average over $\beta$ randomly chosen by the 
flat distribution. 
   Our numerical estimate says that the localization width 
$\mathcal{W}^{(n)}$ in such a case is $n$-independent and is 
given by $\mathcal{W}^{(n)}=\mathcal{W}_{wav}\approx0.736N$ 
approximately. 
   Fig. \ref{loczwidndep}(b) show that some of the localization 
widths are in this range (the broken line interrupted by dots) 
approximately, and they 
actually correspond to the transverse 
Lyapunov mode, namely a wavelike-structure of Lyapunov vectors 
corresponding to the two-point steps of the Lyapunov spectrum 
\cite{Pos00,Tan02c}.


\section{Density Dependence of the Lyapunov Localization}
\label{Locza-Density}

   In this section we compare the 
Lyapunov localization widths $\mathcal{W}^{(n)}$ 
as a function of the Lyapunov index 
at different densities. 
   We concentrate our consideration on the region 
where the Lyapunov spectra change smoothly 
as a function of the Lyapunov index, and show  that 
there are two kinds of Lyapunov index dependence of the 
localization widths: 
exponential dependence [$\mathcal{ED}$] 
and linear dependence [$\mathcal{LD}$]. 
    In the remaining sections of this paper 
the particle number $N=50$ is constant, 
and we change the particle density $\rho$ 
by changing the parameter $d$ connected to $\rho$ 
by Eq. (\ref{densi}).

   
\subsection{Lyapunov Localization at High Density}

   Figure \ref{lyapuwidthhigh} is the Lyapunov localization 
widths $\mathcal{W}^{(n)}/N$ normalized 
by the particle number $N$ as functions of the 
normalized Lyapunov index $n/(2N)$ in the cases of 
$d=-0.05$ (density $\rho=0.826\cdots$, circles), 
$d=0.3$ (density $\rho=0.604\cdots$, triangles) 
and $d=1.5$ (density $\rho=0.314\cdots$, squares). 
   In this figure the localization width 
$\mathcal{W}^{(n)}=\mathcal{W}_{ran}\approx 0.65 N$ 
in the case of random components and the minimum 
localization width $\mathcal{W}^{(n)}=\mathcal{W}_{min}=2$ 
are indicated by the solid line and the broken line, respectively. 

\begin{figure}[!htb]
\vspfigA
\caption{
      Normalized localization widths $\mathcal{W}^{(n)}/N$ 
   as functions of 
   the normalized Lyapunov index $n/(2N)$ 
   in high density cases of $d=-0.05$ (circles), 
   $d=0.3$ (triangles)
   and $d=1.5$ (squares).  
      The data are fitted by exponential functions. 
      The localization widths $\mathcal{W}^{(n)}=\mathcal{W}_{ran}$ 
   and $\mathcal{W}_{min}$ 
   are indicated by the solid line and the broken line, respectively. 
      The localization width indicated by the arrow 
   corresponds to the localized behavior of the Lyapunov vector 
   shown in Fig. \ref{loczat}(a).
   }
\vspfigB
\label{lyapuwidthhigh}\end{figure}  

   In Fig. \ref{lyapuwidthhigh} we fitted the graphs 
to exponential functions 
$y=\alpha_{d}+\beta_{d}\exp(\gamma_{d} x)$ 
with fitting parameters 
$\alpha_{d}$, $\beta_{d}$ and $\gamma_{d}$. 
   Here we find the best values of the fitting parameters to be 
$(\alpha_{-0.05},\beta_{-0.05},\gamma_{-0.05})
=(0.627454,-0.366969,-13.1761)$, 
$(\alpha_{0.3},\beta_{0.3},\gamma_{0.3})
=(0.590977,-0.469104,-9.33407)$ and 
$(\alpha_{1.5},\beta_{1.5},\gamma_{1.5})
=(0.572315,-0.476522,-5.70097)$. 
   The localization widths $\mathcal{W}^{(n)}$ 
are nicely fitted by these 
exponential functions in the region $n/(2N) \leq 0.6$ 
(the exponential dependence [$\mathcal{ED}$]). 

   Figure \ref{lyapuwidthhigh} also shows that the localization 
widths decrease as the parameter $d$ increases 
(therefore as the density $\rho$ decreases). 
   The largest Lyapunov exponent is an increasing function of 
the particle density, as will be discussed 
in Sect. \ref{densitydependenceLmax}, 
so this result means that the Lyapunov 
localization is stronger in a system with weaker chaos characterized by 
a smaller value of the largest-Lyapunov exponent.
   It should also be noted that in this region the localization 
widths are smaller than the localization width of random elements, 
which is about $\mathcal{W}_{ran}\approx 0.65 N$, and 
larger than $\mathcal{W}_{min} = 0.04 N$. 
   Fig. \ref{loczat}(a) in Sect. \ref{LocalizationWidth} was 
the behavior of the Lyapunov localization corresponding 
to the localization width indicated by the arrow in 
Fig. \ref{lyapuwidthhigh}.
   It is important to note that the shape of the quantities 
$\gamma_{j}^{(n)}(t)$ as a function of the particle index $j$ 
have a tail like that in Fig. \ref{loczat}(a) in the region 
showing an exponential  
dependence [$\mathcal{ED}$] of the localization widths.



\subsection{Lyapunov Localization at Low Density}
\label{loczatinLowDens}

   Now we consider the Lyapunov localization in low density cases, 
given by large values of the parameter $d$.

\begin{figure}[!htb]
\vspfigA
\caption{
      Normalized localization widths $\mathcal{W}^{(n)}/N$ 
   as functions of the normalized Lyapunov index $n/(2N)$ 
   in low density cases of 
   $d=10^{2}$ (squares), 
   $d=5\times10^{2}$ (triangles) and 
   $d=10^{5}$ (circles). 
      The localization widths $\mathcal{W}^{(n)}=\mathcal{W}_{ran}$ and 
   $\mathcal{W}_{min}$ 
   are indicated by the solid line and the broken line, respectively. 
      (a) Localization widths in the Lyapunov index region 
   $n/(2N) \leq 0.7$. 
      The data are fitted by an exponential function.  
      The grey region is the region in which the localization 
   widths show a linear dependence [$\mathcal{LD}$] 
   as a function of Lyapunov index 
   in the low density limit. 
      (b) Enlarged graphs including a linear dependence region 
   of the localization widths. 
       The data are fitted by linear functions. 
      This figure also includes a part of the graph of the 
   normalized localization widths 
   in the case of $d=1.5$ (the crosses), 
   which has already been shown in 
   Fig. \ref{lyapuwidthhigh}.   
      The localization widths indicated by the black arrow,  
   the grey arrow and the outline arrow correspond 
   to localized behaviors of the Lyapunov vectors shown 
   in Figs. \ref{loczat}(a), \ref{loczat}(b) and \ref{loczat2}, 
   respectively. 
      The thick grey horizontal line to connect 
   the localization widths indicated by the black arrow 
   and the outline arrow in Fig. (b) 
   is given to show that these values 
   take almost the same value. 
   }
\vspfigB
\label{lyapuwidthlow}\end{figure}  

   Figure \ref{lyapuwidthlow}(a) is the 
Lyapunov localization widths $\mathcal{W}^{(n)}/N$ 
normalized by the particle number $N$ 
as functions of normalized Lyapunov index $n/(2N)$ 
in the case of 
$d=10^{2}$ (density $\rho=0.00777\cdots$, squares), 
$d=5\times10^{2}$ (density $\rho=0.00156\cdots$, triangles) 
and $d=10^{5}$ (density $\rho=0.00000785\cdots$, circles), 
which correspond to the region where the Lyapunov spectra 
change smoothly.
   It is clear that the Lyapunov index dependences of the 
localization widths in this figure are different from the 
high density cases shown in Fig. \ref{lyapuwidthhigh} in 
a couple of senses. 
   First, linear dependences of the localization widths with 
respect to Lyapunov index appear in the region of small 
localization widths, which correspond to large 
Lyapunov exponents. 
   In Fig. \ref{lyapuwidthlow}(b), fits of Lyapunov 
index dependences 
of the localization widths by linear functions 
$y=\tilde{\alpha}_{d} x + \tilde{\beta}_{d}$ are given 
with fitting parameters $\tilde{\alpha}_{d}$ 
and $\tilde{\beta}_{d}$: 
$(\tilde{\alpha}_{10^{2}}, \tilde{\beta}_{10^{2}}) 
= (0.503874,0.046405)$ in the case of $d=10^{2}$, 
$(\tilde{\alpha}_{5\times10^{2}}, \tilde{\beta}_{5\times10^{2}}) 
= (0.347437,0.044468)$ in the case of $d=5\times10^{2}$, 
and $(\tilde{\alpha}_{10^{5}}, \tilde{\beta}_{10^{5}})  
= (0.224129,0.0427506)$ in the case of $d=10^{5}$. 
   It is important to note that the localization widths are always 
larger than $2$, namely $\mathcal{W}^{(n)}/N > 2/N = 0.04$ 
in this figure, and the smallest localization widths 
corresponding to the largest Lyapunov exponents are close 
to this minimum value in these low density cases.  
   Besides, graphs of the linear dependence of the localization 
widths is flatter in the lower density case, 
as shown by the fact that the 
value of the fitting parameter $\tilde{\alpha}_{d}$ 
decreases as the value of the parameter $d$ increases.  
   The existence of the linear dependence [$\mathcal{LD}$] 
of the localization widths is one of the main results 
of this paper. 
   (In Fig. \ref{lyapuwidthlow}(a) the region of 
the linear dependence [$\mathcal{LD}$] in the low density limit 
is shaded grey.)
   Secondly, in Fig. \ref{lyapuwidthlow}(a), 
we can still recognize a region 
of localization widths in which  the Lyapunov index dependence 
of the Lyapunov localization widths is exponential 
(the dependence [$\mathcal{ED}$]). 
   To show it clearly we give a fit of an exponential function 
$y=\alpha'+\beta'\exp(\gamma' x)$ 
with values 
$(\alpha',\beta',\gamma') = (0.560814, -2.73348, -7.50543)$ 
of the fitting parameters in Fig. \ref{lyapuwidthlow}(a). 
   It should be emphasized that 
the shapes of localization widths 
are almost independent of the particle density $\rho$, 
at least in the three density cases shown 
in Fig. \ref{lyapuwidthlow}(a). 
   As the third point one may notice that from above characteristics 
of the localization widths in the low density cases the region of 
the linear dependence [$\mathcal{LD}$] and the region of 
the exponential dependence [$\mathcal{ED}$] 
are more distinguishable in a lower density case, 
with an accompanying sharp bending of the localization width profile. 
   Figure \ref{lyapuwidthlow} suggests that the spectrum of the 
localization widths as a function of Lyapunov index 
has a distinct shape in the low density limit 
$\rho\rightarrow 0$ with a critical value $\zeta_{EL}$ of the 
Lyapunov index where the localization width 
$\mathcal{W}^{(n)}$ shows a linear dependence 
[$\mathcal{LD}$] in the Lyapunov index 
$n\leq\zeta_{EL}$ and 
shows an exponential dependence 
[$\mathcal{ED}$] in the Lyapunov index $n>\zeta_{EL}$. 

\begin{figure}[!htb]
\vspfigA
\caption{
      Localized behaviors of the amplitudes 
   $\gamma_{j}^{(15)}$ 
   of the normalized Lyapunov vector components of each particle 
   as a function of the collision number 
   $n_{t}$ and the particle index $j$ in the case of 
   $d=5\times10^{2}$. 
   The corresponding localization width is indicated by 
   the outline arrow in Fig. \ref{lyapuwidthlow}(b).}
\vspfigB
\label{loczat2}\end{figure}  

   It is important to note that the exponential dependence 
[$\mathcal{ED}$] and the linear dependence [$\mathcal{LD}$] 
of the localization widths can also be distinguished by the shapes 
of the amplitudes $\gamma_{j}^{(n)}(t)$ of the normalized 
Lyapunov vector components of each particle. 
   To discuss this point it is important to note 
that Fig. \ref{loczat}(a) is the  
localized behavior of the Lyapunov vector 
corresponding to the Lyapunov width 
indicated by the black arrow in Figs. \ref{lyapuwidthhigh} 
and \ref{lyapuwidthlow}(b)
and is in the region 
of exponential dependence [$\mathcal{ED}$] of 
localization widths. 
   This localization shows a long tail behavior. 
   On the other hand Fig. \ref{loczat}(b) is a Lyapunov 
localization corresponding to the localization width indicated 
by the grey arrow in Fig. \ref{lyapuwidthlow}(b) 
and is in a region of linear dependence [$\mathcal{LD}$] 
of localization widths. 
   This localization shows a sharp rectangular shape 
(with width $2$). 
   These observations lead to the conjecture 
that in the exponential dependence [$\mathcal{ED}$] 
the amplitudes $\gamma_{j}^{(n)}(t)$ have a long tail, 
and in the linear dependence [$\mathcal{LD}$] 
the amplitudes $\gamma_{j}^{(n)}(t)$ 
have a rather sharp rectangular shape. 
   To make this point convincing, we compare the behaviors of 
the amplitudes $\gamma_{j}^{(n)}(t)$ corresponding to two 
localization widths which take almost the same value of the 
localization widths but show different 
dependences [$\mathcal{LD}$] and [$\mathcal{ED}$] of 
the localization widths. 
   For this purpose we choose the $15$-th localization width 
$\mathcal{W}^{(15)}$ for $d=5\times10^{2}$. 
   This localization width, indicated by the outlined 
arrow in Fig. \ref{loczat}(b),
is in the region of the linear dependence [$\mathcal{LD}$] and 
takes almost the same value as the localization width 
$\mathcal{W}^{(1)}$ indicated by the black arrow in 
Figs. \ref{lyapuwidthhigh} 
and \ref{lyapuwidthlow}(b) in the case $d=1.5$, which is 
in the region of exponential dependence [$\mathcal{ED}$] of the 
localization widths, and 
whose localized behavior is shown in Fig. \ref{loczat}(a). 
   Figure \ref{loczat2} is the localized behavior 
of the amplitudes $\gamma_{j}^{(15)}(t)$ as a function 
of the collision number $n_{t}$ and the particle index $j$ 
in the case $d=5\times10^{2}$.  
   Here, the data in Fig. \ref{loczat2} is taken after every
second collision. 
   This figure shows that some peaks of the amplitudes 
$\gamma_{j}^{(n)}(t)$ have flat tops with a width $2$ 
and distinct rectangular shapes, 
although its corresponding Lyapunov exponent, the 15-th Lyapunov 
exponent is far from the largest one and it is less 
localized than in Fig. \ref{loczat}. 
 
   It should be noted that the linear dependence 
[$\mathcal{LD}$] of the localization widths as a function of 
Lyapunov index is not a property 
of the quasi-one-dimensional systems only. 
   In Appendix \ref{AppendixLocalizationWidths} we 
consider the localization widths in a square system, 
and show that the linear linear dependence [$\mathcal{LD}$] 
of the Lyapunov localization also appears in the square system.


\section{Lyapunov Spectra, and the Angles and Amplitudes of Lyapunov Vector Components}
\label{spectraandangles}

   In this section, we consider how differences between the linear 
[$\mathcal{LD}$] and exponential dependences 
[$\mathcal{ED}$] of the localization widths affect
other quantities, such as the Lyapunov spectrum, 
the angle between the spatial and 
momentum parts of the Lyapunov vectors, and the amplitudes of  the 
$x$ and $y$-components of the spatial part 
and the momentum part of the Lyapunov vectors. 
   We also investigate the effects of the stepwise structure of the 
Lyapunov spectra on these quantities.


\subsection{Lyapunov Spectra}
\label{Lyapunovspectra}

   Figure \ref{lyaspe} is the Lyapunov spectra normalized by 
their largest Lyapunov exponents $\lambda^{(1)}$ for the systems 
as functions of the normalized Lyapunov index $n/(2N)$ 
for 
$d=-5\times10^{-2}$ (white-filled circles), 
$d=0.3$ (white-filled triangles), 
$d=1.5$ (white-filled squares), 
$d=10^{2}$  (black-filled squares), 
$d=5\times 10^{2}$  (black-filled triangles) 
and $d=10^{5}$  (black-filled circles). 
   Here the values of the largest Lyapunov exponents are 
given by 
$\lambda^{(1)} \approx 4.79$ for $d=-5\times10^{-2}$, 
$\lambda^{(1)} \approx 1.64$ for $d=0.3$, 
$\lambda^{(1)} \approx 0.769$ for $d=1.5$, 
$\lambda^{(1)} \approx 0.0579$ for $d=10^{2}$, 
$\lambda^{(1)} \approx 0.0161$ for $d=5\times 10^{2}$, 
$\lambda^{(1)} \approx 0.000157$ for $d=10^{5}$. 
   The negative branch of the Lyapunov spectra are omitted 
in this figure, because of the conjugate property 
(\ref{Sympl3}) of the Lyapunov exponents. 
   The Lyapunov index dependence of the localization widths 
in these cases were shown in Figs. 
\ref{lyapuwidthhigh} and \ref{lyapuwidthlow}. 
   It should be emphasized that in the low density cases, 
in which the linear dependence [$\mathcal{LD}$] appears, 
we can recognize a sharp bend in the Lyapunov spectra around 
the normalized Lyapunov index $n = \zeta_{EL}$. 
   Such a bending point corresponds to the point that distinguishes 
the dependences [$\mathcal{LD}$] and [$\mathcal{ED}$] 
of the localization widths in Fig. \ref{lyapuwidthlow}. 
   (In this figure the region of the linear dependence 
[$\mathcal{LD}$] of localization widths at low densities 
has a grey background.)
   It may be noted that such a bending of the Lyapunov 
spectrum has also been reported in a fully two-dimensional system 
\cite{Gas02note} 
(Also see the Lyapunov spectra for the Fermi-Pasta-Ulum models 
in Refs. \cite{Liv87b,yam98}.). 

\begin{figure}[!htb]
\vspfigA
\caption{
      Lyapunov spectra normalized by the largest Lyapunov 
   exponents $\lambda^{(1)}$ 
   as functions of normalized Lyapunov index $n/(2N)$ 
   in the cases of 
   $d=-5\times10^{-2}$ (white-filled circles), 
   $d=0.3$ (white-filled triangles), 
   $d=1.5$ (white-filled squares), 
   $d=10^{2}$  (black-filled squares), 
   $d=5\times 10^{2}$  (black-filled triangles) 
   and $d=10^{5}$  (black-filled circles). 
      The grey region is the region in which the localization 
   widths show a linear dependence [$\mathcal{LD}$] 
   as a function of Lyapunov index at low densities. 
      The small figure in the upper right side is the enlarged 
    graphs of the small positive region of the Lyapunov spectra 
    for the low density cases of $d=10^{2}$, 
    $5\times 10^{2}$  and $10^{5}$. 
   }
\vspfigB
\label{lyaspe}\end{figure}  

   In Fig. \ref{lyaspe} we can see some stepwise structures 
of the Lyapunov spectra, not only in the high density cases 
$d=-5\times10^{-2}$, $d=0.3$ and $1.5$, 
but also in the low density cases $d=10^{2}$, 
$d=5\times 10^{2}$  and $d=10^{5}$.  
   (See the enlarged graphs in the 
upper right side of Fig. \ref{lyaspe} for the stepwise structure 
of the Lyapunov spectra in these low density cases.) 
   The steps of the Lyapunov spectra consist of two-point steps 
and four point steps, which were considered in Ref. \cite{Tan02c} 
in detail. 
   Different from the high density cases, at low density the separations of the 
two-point steps and the four-point steps 
in the Lyapunov spectra are not so clear.


\subsection{Angle Between the Spatial and the Momentum Parts of the Lyapunov Vectors}
\label{Anglebetweenthepoatial}

   As a second example to illustrate the effects of 
the two dependences [$\mathcal{LD}$] 
and [$\mathcal{ED}$] of localization widths, 
we consider the angle $\theta^{(n)}$ between the 
spatial part $\delta\mathbf{q}^{(n)} 
= (\delta\mathbf{q}_{1}^{(n)},\delta\mathbf{q}_{2}^{(n)},
\cdots,\delta\mathbf{q}_{N}^{(n)} )^{T}$ 
and the momentum  
part $\delta\mathbf{p}^{(n)} 
= (\delta\mathbf{p}_{1}^{(n)},\delta\mathbf{p}_{2}^{(n)},
\cdots,\delta\mathbf{p}_{N}^{(n)} )^{T}$ 
of the Lyapunov vector $\delta\bfGamma^{(n)}$ 
corresponding to the $n$-th Lyapunov exponent $\lambda^{(n)}$, 
which is defined by  

\begin{eqnarray}
   \theta^{(n)}\equiv 
		   \cos^{-1}\left(\frac{
		      \delta \mathbf{q}^{(n)} \cdot \delta \mathbf{p}^{(n)} }
		      {\left|\delta \mathbf{q}^{(n)}\right|
      \left|\delta \mathbf{p}^{(n)}\right|
		   }\right) .
\label{angleLyapuvecto}\end{eqnarray}

\noindent Noting the conjugate relations  

\begin{eqnarray}
		   \frac{
      \delta \mathbf{q}^{(4N-n+1)} \cdot \delta \mathbf{p}^{(4N-n+1)}}
		      {\left|\delta \mathbf{q}^{(4N-n+1)}\right|
      \left|\delta \mathbf{p}^{(4N-n+1)}\right|}
		   = - \frac{
      \delta \mathbf{q}^{(n)} \cdot \delta \mathbf{p}^{(n)}}
		      {\left|\delta \mathbf{q}^{(n)}\right|
      \left|\delta \mathbf{p}^{(n)}\right|}
\end{eqnarray}

\noindent following from the conjugate property (\ref{Sympl}) 
of the Lyapunov vectors,  
the angle $\theta^{(n)}$ satisfies the condition

\begin{eqnarray}
   \theta^{(4N-n+1)} = \pi - \theta^{(n)}.
\label{anglesymple}\end{eqnarray}

\noindent This angle has already been investigated 
in many-particle systems \cite{Tan02c,Mcn01b}. 
   It should be noted that some analytical approaches 
give formula to calculate Lyapunov exponents from 
the spatial part only 
(or the momentum part only) of 
the tangent vector \cite{Tan02a,Bei98,Zon98}, 
so it is important 
to investigate the relation between the 
spatial part and the momentum part 
of the Lyapunov vector.

\begin{figure}[!htb]
\vspfigA
\caption{
      Time-average $\langle\theta^{(n)}\rangle/\pi$ of the angle 
   divided by $\pi$ for 
   the spatial part $\delta \mathbf{q}^{(n)}$ and the 
   momentum part $\delta \mathbf{p}^{(n)}$ of the Lyapunov vector 
   corresponding to the $n$-th Lyapunov exponent 
   $\lambda^{(n)}$ as functions of the normalized 
   Lyapunov index $n/(2N)$ for 
   $d=-5\times10^{-2}$ (white-filled circles), 
   $d=0.3$ (white-filled triangles), 
   $d=1.5$ (white-filled squares), 
   $d=10^{2}$  (black-filled squares), 
   $d=5\times 10^{2}$  (black-filled triangles) 
   and $d=10^{5}$  (black-filled circles).
      The grey region is the region in which the localization widths 
   show a linear dependence [$\mathcal{LD}$] 
   as a function of Lyapunov index in low density cases. 
      The solid line corresponds to 
   the value $\theta^{(n)}=\pi/2$ of the angle.
      The small figure in the upper right side is the enlarged 
    graphs for the low density cases $d=10^{2}$, 
    $5\times 10^{2}$ and $10^{5}$ 
    in the region of large Lyapunov indices. }
\vspfigB
\label{anglyavec}\end{figure}  

   In Fig. \ref{anglyavec} we show the graphs of the time-average 
$\langle\theta^{(n)}\rangle/\pi$ of the angle 
$\theta^{(n)}$ divided by $\pi$ 
as functions of the normalized Lyapunov index $n/(2N)$ in the cases 
$d=-5\times10^{-2}$ (white-filled circles), 
$d=0.3$ (white-filled triangles), 
$d=1.5$ (white-filled squares), 
$d=10^{2}$  (black-filled squares), 
$d=5\times 10^{2}$  (black-filled triangles) 
and $d=10^{5}$  (black-filled circles). 
   Here the time-average of the angle $\theta^{(n)}$ is the 
arithmetic average of their values immediately after collision, 
for $1000N$ collisions. 
   The plots of the time-averaged angles 
$\langle\theta^{(n)}\rangle$, $n=2N+1,2N+2,\cdots,4N$, 
corresponding to negative branch of the Lyapunov exponents 
are omitted because of the conjugate property 
(\ref{anglesymple}) 
of the angle $\theta^{(n)}$. 
   It is noted that in our numerical calculations 
the amplitudes $|\delta \mathbf{p}^{(n)}|$, $n=2N-2$, $2N-1$ and 
$2N$ corresponding to zero-Lyapunov exponents are 
zero (or extremely small), 
so the angles $\theta^{(n)}$, $n=2N-2$, $2N-1$ and 
$2N$ cannot be defined. 
   This is the reason why we do not 
plot these angles 
in Fig.  \ref{anglyavec} (See the subsection 
\ref{amplitudecomponents} about this point.).
   This figure shows that in the low density cases, 
in which the linear dependence [$\mathcal{LD}$] 
of the localization widths appears, the 
spectra of the averaged angle 
$\langle\theta^{(n)}\rangle$ bends around the value
$n/(2N) = \zeta_{EL}/(2N)$ of the normalized Lyapunov 
index. 
   (In Fig. \ref{anglyavec} we showed a region of 
the linear dependence [$\mathcal{LD}$] of the 
localization widths as a grey region.)
   In the region of the linear dependence 
[$\mathcal{LD}$] of the localization widths 
the time-averaged angles 
$\langle\theta^{(n)}\rangle$ increase 
monotonically as functions of Lyapunov indices $n$, 
while the time-averaged angles 
$\langle\theta^{(n)}\rangle$ corresponding 
to the exponential dependence [$\mathcal{ED}$] 
of the localization widths is close to 
$\pi/2$, meaning that in this region the vector 
$\delta \mathbf{q}^{(n)}$ is almost orthogonal to the vector  
$\delta \mathbf{p}^{(n)}$ (See the case of $d=10^{5}$ in Fig. 
\ref{anglyavec}.). 

   We can also see the effect of the stepwise structure of the 
Lyapunov spectra in the angle $\theta^{(n)}$. 
   As shown in Fig. \ref{anglyavec} the values of the time-averaged 
angles $\langle\theta^{(n)}\rangle$ corresponding to the 
two-point steps of the Lyapunov spectra in Fig. \ref{lyaspe} are 
rather smaller than the values corresponding to their four-point 
steps, in all densities shown in Fig. \ref{anglyavec} 
(See the upper right side of Fig. 
\ref{anglyavec} for the low density cases). 
   Therefore the Lyapunov index dependence of the time-averaged 
angles $\langle\theta^{(n)}\rangle$ can be used to distinguish 
the two-point step of the Lyapunov spectra from the other parts.


\subsection{Amplitudes of the $x$ and $y$-Components of the Lyapunov Vectors}
\label{amplitudecomponents}

   As shown in the subsection \ref{Lyapunovspectra}, in low 
density cases  
the positive branch of Lyapunov spectra 
is separated into three parts: 
(i) A region in which the Lyapunov spectrum is a 
rapidly decreasing function of Lyapunov index and 
corresponds to the linear dependence [$\mathcal{LD}$] of 
the localization widths,  
(ii) A region in which values of the Lyapunov exponents are 
very small compared to the largest Lyapunov exponent and 
corresponds to the exponential dependence [$\mathcal{ED}$] 
of the localization widths, and 
(iii) A region showing the stepwise structure 
of the Lyapunov spectra. 
   The boundary of the regions (i) and (ii) of the Lyapunov 
spectra appear as a sharp bending of the Lyapunov spectra 
in the low density limit. 
   To understand this characteristic of the Lyapunov spectra 
we note that a positive Lyapunov exponent is connected to 
a decorrelation rate in chaotic dynamics, 
and a larger positive Lyapunov exponent means  
that a trajectory diverges more rapidly on the energy surface 
in ergodic systems and loses correlation with 
its initial value more quickly. 
   This is partly supported by the fact that 
the Kolmogorov-Sinai entropy, which is equal to the sum of 
all the positive Lyapunov exponents (Pesin's identity) 
in closed systems, 
is connected to the inverse relaxation time 
in ergodic systems \cite{Gas98}. 
   A relation between a Lyapunov exponent and a decay rate 
of time-correlation is also discussed in Refs. \cite{Col03,Del97}. 
   Another example of a relation between a time-scale of the 
 dynamics and a positive Lyapunov exponent is the 
tracer particle effect in Lyapunov spectra \cite{Gas02}. 
   On this subject a many-particle system consisting of a small 
light particle called the tracer particle moving at high speed 
(the short time scale) and many 
big heavy particles moving at low speeds 
(the long time scale) is considered, 
and it is shown that the existence of the tracer particle 
leads to relatively large Lyapunov exponents compared with 
the Lyapunov exponents corresponding to the other heavy 
particles. 
   The above considerations suggest that 
at low density the quasi-one-dimensional systems 
exhibit a bending of the Lyapunov spectra, 
which separates the set of positive Lyapunov exponents 
into two groups (i) and (ii) of relatively different values,   
and this suggests the existence of a time-scale 
separation in the dynamics of these systems. 

   One might think that in the quasi-one-dimensional 
systems such a time-scale separation 
may come from the very narrow rectangular shape of the system. 
   In the quasi-one-dimensional system, especially 
in the large $d$ case with a low density, the time-scale of 
oscillation of each particle in the $x$-direction is 
much larger than the time-scale of that in the $y$-direction, 
and it may be one possibility leading to the time-scale separation 
mentioned above. 
   (However this possibility may be rather unlikely, 
because as shown in Appendix \ref{AppendixLyapunovSpectra} 
the bending point of the Lyapunov spectra does not change 
significantly by changing the system from the quasi-one-dimensional 
system to the square system with the same area 
$L_{x}L_{y}$ and the 
same number of particles $N$.)
   Another possibility to explain this bending 
of the Lyapunov spectrum may be from 
the different roles of the spatial 
and momentum parts of the Lyapunov vectors 
at low density. 
   The result in the previous subsection 
supports this point, showing that the spatial 
and momentum parts of the Lyapunov vectors 
are not in the same direction at low density.

   Motivated by the above considerations, in this subsection 
we consider the amplitudes of the four parts of the 
Lyapunov vectors: the $x$ and 
$y$-components in the spatial 
part and the momentum part of the Lyapunov vectors, 
namely the four kinds of quantities 
defined by

\begin{eqnarray}
   \eta_{qj}^{(n)}(t) \equiv 
   \frac{\sum_{k=1}^{N}\left| \delta q_{jk}^{(n)}(t) \right|^{2}}
   {\sum_{k=1}^{N} 
   \left| \delta\bfGamma_{k}^{(n)}(t) \right|^{2}} 
\label{lyaamp2}\end{eqnarray}
\begin{eqnarray}
   \eta_{pj}^{(n)}(t) \equiv 
   \frac{\sum_{k=1}^{N}\left| \delta p_{jk}^{(n)}(t) \right|^{2}}
   {\sum_{k=1}^{N} 
   \left| \delta\bfGamma_{k}^{(n)}(t) \right|^{2}} 
\label{lyaamp3}\end{eqnarray}

\noindent $j=x$ and $y$, in each Lyapunov index $n$. 
   Here $\delta q_{jk}^{(n)}(t)$ and $\delta p_{jk}^{(n)}(t)$ 
are the $j$-component of the spatial coordinate part 
$\delta \mathbf{q}_{k}^{(n)} 
= (\delta q_{xk}^{(n)}(t),\delta q_{yk}^{(n)}(t))$ 
and the momentum part 
$\delta \mathbf{p}_{k}^{(n)} 
= (\delta p_{xk}^{(n)}(t),\delta p_{yk}^{(n)}(t))$ 
of the Lyapunov vector 
$\delta \bfGamma_{k}^{(n)} 
=(\delta \mathbf{q}_{k}^{(n)},\delta \mathbf{p}_{k}^{(n)})$
corresponding to the $k$-th particle and the 
$n$-th Lyapunov exponent at time $t$, 
respectively. 
   These four quantities are normalized as 

\begin{eqnarray}
   \eta_{px}^{(n)}(t) + \eta_{py}^{(n)}(t) 
   + \eta_{qx}^{(n)}(t) + \eta_{qy}^{(n)}(t) = 1 
\label{lyaamp23nor}\end{eqnarray}

\noindent by their definitions. 
   Using the conjugate property (\ref{Sympl}) 
of the Lyapunov vector we obtain 

\begin{eqnarray}
   \eta_{pj}^{(4N-n+1)}(t) = \eta_{qj}^{(n)}(t)\\
   \eta_{qj}^{(4N-n+1)}(t) = \eta_{pj}^{(n)}(t).
\label{lyaamp4}\end{eqnarray}

\noindent Therefore we can omit 
the quantities 
$\eta_{pj}^{(4N-n+1)}(t)$ and $\eta_{qj}^{(4N-n+1)}(t)$, 
for $n=1,2,\cdots,2N$ corresponding to 
the negative branch of the Lyapunov spectra.

\begin{figure}[!htb]
\vspfigA
\caption{
      Time-averages of the normalized amplitudes of 
   the $x$-component of the spatial part 
   ($\langle\eta_{qx}^{(n)}(t)\rangle$, circles), 
   the $y$-component of the spatial coordinate part 
   ($\langle\eta_{qy}^{(n)}(t)\rangle$, triangles), 
   the $x$-component of the momentum part
   ($\langle\eta_{px}^{(n)}(t)\rangle$, crosses), 
   and its $y$-component of the momentum part 
   ($\langle\eta_{py}^{(n)}(t)\rangle$, diagonal crosses) 
   of the Lyapunov vectors 
   as functions of the normalized Lyapunov index $n/(2N)$ 
for 
   $d=-5\times 10^{-2}$ (the graph (a)), 
   $d=0.1$ (the graph (b)), 
   $d=1$ (the graph (c)) and 
   $d=10^{5}$ (the graph (d)). 
      The small panels in the top right of figures 
   (a), (b), (c) and (d) 
   are enlarged graphs of $\langle\eta_{px}^{(n)}(t)\rangle$ 
   and $\langle\eta_{py}^{(n)}(t)\rangle$ 
   in small positive Lyapunov exponent regions. 
      The grey region in the graph (d) 
   is the region in which the localization widths 
   show a linear dependence [$\mathcal{LD}$] 
   as a function of Lyapunov index. 
   }
\vspfigB
\label{lyaveccomp}\end{figure}  

   Figure \ref{lyaveccomp} is the graphs of the time-averaged 
quantities 
$\langle\eta_{qx}^{(n)}(t)\rangle$ (circles), 
$\langle\eta_{qy}^{(n)}(t)\rangle$ (triangles), 
$\langle\eta_{px}^{(n)}(t)\rangle$ (crosses) and 
$\langle\eta_{py}^{(n)}(t)\rangle$ (diagonal crosses) of 
the quantities defined by Eqs. (\ref{lyaamp2}) and 
(\ref{lyaamp3}) as functions of the normalized 
Lyapunov index $n/(2N)$  for 
$d=-5\times 10^{-2}$ (Fig. \ref{lyaveccomp}(a), 
density $\rho=0.826\cdots$), 
$d=0.1$ (Fig. \ref{lyaveccomp}(b), 
density $\rho=0.713\cdots$), 
$d=1$ (Fig. \ref{lyaveccomp}(c), 
density $\rho=0.392\cdots$) and 
$d=10^{5}$ (Fig. \ref{lyaveccomp}(d), 
density $\rho=0.00000785\cdots$). 
   Here the time-average $\langle\mathcal{X}\rangle$ 
 of the quantities $\mathcal{X} = $
$\eta_{qx}^{(n)}(t)$, 
$\eta_{qy}^{(n)}(t)$
$\eta_{px}^{(n)}(t)$ and 
$\eta_{py}^{(n)}(t)$ is taken as 
the arithmetic average of the quantity $\mathcal{X}$ immediately 
after collisions, for $1000N$ collisions. 
   The cases of $d=-5\times 10^{-2}$ and $d=10^{5}$ 
shown in Figs. \ref{lyaveccomp}(a) and \ref{lyaveccomp}(d) 
are the cases 
of the highest density and the lowest density considered 
in Sect. \ref{Locza-Density} and the previous two subsections 
of this section, respectively, and the other two cases 
are given to show intermediate situations between 
the two cases of Figs. \ref{lyaveccomp}(a) and \ref{lyaveccomp}(d).
   It is noted that Figs. \ref{lyaveccomp}(a), (b) 
and (c) are for the case of a particle density 
in which the linear dependences [$\mathcal{LD}$] 
of localization widths do not appear yet, and 
only in the grey region of Figs. \ref{lyaveccomp}(d) 
does the linear dependence [$\mathcal{LD}$] appear. 
   From this figure it is clear 
that there is a strong asymmetry between 
the spatial and momentum parts of the 
Lyapunov vectors in any density region, as well as 
an asymmetry between the $x$ and $y$-components 
of the Lyapunov vectors especially at low density. 

   The behavior of the graphs in Fig. \ref{lyaveccomp} 
are different in the region where the Lyapunov spectra 
change smoothly and in the stepwise region of the 
Lyapunov spectra. 
   First we discuss the region where the Lyapunov 
spectra change smoothly, which is the region of the 
Lyapunov indexes $n/(2N) < 0.8$ approximately 
in Fig. \ref{lyaveccomp}. 
   In the high density case shown in Fig. \ref{lyaveccomp}(a) 
the momentum parts ($\langle\eta_{px}^{(n)}(t)\rangle$ and 
$\langle\eta_{py}^{(n)}(t)\rangle$) of the Lyapunov vectors 
are larger than the corresponding spatial parts 
($\langle\eta_{qx}^{(n)}(t)\rangle$ and  
$\langle\eta_{qy}^{(n)}(t)\rangle$).  
   As the density decreases, the spatial parts 
of the Lyapunov vectors increase (and momentum parts decrease), 
first in the \emph{large} positive Lyapunov exponent region 
(See Fig. \ref{lyaveccomp}(a) $\rightarrow$ (b).), and 
then in the \emph{small} Lyapunov exponent region 
(See Fig. \ref{lyaveccomp}(b) $\rightarrow$ (c) $\rightarrow$ (d).).
  At very low density, as shown in Fig. \ref{lyaveccomp}(d), 
the spatial parts of the Lyapunov vectors are much 
larger than the corresponding momentum parts 
for Lyapunov vectors of any index,
and a linear dependence [$\mathcal{LD}$] of the  
localization widths appears. 
   In this density region a significant gap in the $x$
and $y$-components of the spatial part of the Lyapunov vectors 
appears, and the transverse part 
($\langle\eta_{qy}^{(n)}(t)\rangle$) 
is larger than the longitudinal part 
$\langle\eta_{qx}^{(n)}(t)\rangle$ 
in the large positive Lyapunov exponent region while 
while the reverse is true in the small positive Lyapunov exponent region. 
   In the low density limit the momentum parts 
$\langle\eta_{px}^{(n)}(t)\rangle$ and  
$\langle\eta_{py}^{(n)}(t)\rangle$ are extremely small 
so the spatial parts $\langle\eta_{qx}^{(n)}(t)\rangle$ and  
$\langle\eta_{qy}^{(n)}(t)\rangle$ are almost symmetric 
in the line $y=0.5$ as required by the the normalization condition 
(\ref{lyaamp23nor}).
   It should be emphasized that
in the Lyapunov index of the linear dependence 
[$\mathcal{LD}$] of the localization widths the quantities 
$\langle\eta_{qy}^{(n)}(t)\rangle$ and  
$\langle\eta_{qx}^{(n)}(t)\rangle$ are rather flat, 
shown as the grey region in Fig. \ref{lyaveccomp}(d). 
   We can show that these asymmetries between the 
$x$ and $y$-components 
of the Lyapunov vectors actually come from 
its narrow rectangular shape, because 
as shown in 
Appendix \ref{AppendixAmplitudesofXcomponents}
such asymmetries do not appear 
in the square system, although the asymmetries 
between the spatial and 
momentum parts of the Lyapunov vectors still 
exist there. 

   As a second point, Fig. \ref{lyaveccomp} also shows 
some characteristics of the time-averaged amplitudes of 
the Lyapunov vector components corresponding to the stepwise 
region of the Lyapunov spectra. 
   In Fig. \ref{lyaveccomp} we can see flat parts 
consisting of two-points (four-points) for the 
time-averaged normalized amplitudes 
$\langle\eta_{qx}^{(n)}(t)\rangle$, 
$\langle\eta_{qy}^{(n)}(t)\rangle$, 
$\langle\eta_{px}^{(n)}(t)\rangle$ and 
$\langle\eta_{py}^{(n)}(t)\rangle$ of the Lyapunov vectors,  
which are corresponding to the two-point steps 
(the four-point steps) of the Lyapunov spectra  
(See Figs. \ref{lyaveccomp}(a), (b), (c) and (d) in 
a region of the Lyapunov index around $n/(2N) > 0.9$, 
as well as the small panels in the top right of these 
figures for enlarged momentum parts 
$\langle\eta_{px}^{(n)}(t)\rangle$ and 
$\langle\eta_{py}^{(n)}(t)\rangle$ in the 
small positive Lyapunov exponent regions.). 
   Values of the two-point steps of the transverse part 
$\langle\eta_{qy}^{(n)}(t)\rangle$ and 
$\langle\eta_{py}^{(n)}(t)\rangle$ of the Lyapunov vectors 
corresponding to the two-point steps of the Lyapunov spectra 
are larger than their longitudinal parts 
$\langle\eta_{qx}^{(n)}(t)\rangle$ and 
$\langle\eta_{px}^{(n)}(t)\rangle$ 
in Figs. \ref{lyaveccomp}(a), (b), (c) and (d). 
   On the other hand, values of the four-point steps 
of the longitudinal part 
$\langle\eta_{qx}^{(n)}(t)\rangle$ and 
$\langle\eta_{px}^{(n)}(t)\rangle$ of the Lyapunov vectors 
corresponding to the four-point steps of the Lyapunov spectra 
are larger than their transverse parts 
$\langle\eta_{qy}^{(n)}(t)\rangle$ and 
$\langle\eta_{py}^{(n)}(t)\rangle$ 
in Figs. \ref{lyaveccomp}(a), (b) and (c), 
and the spatial parts of the Lyapunov vectors 
in Figs. \ref{lyaveccomp}(d). 
   At any density the spatial parts 
$\langle\eta_{qx}^{(n)}(t)\rangle$ and  
$\langle\eta_{qy}^{(n)}(t)\rangle$ are rather larger  
than the momentum parts  
$\langle\eta_{px}^{(n)}(t)\rangle$ and  
$\langle\eta_{py}^{(n)}(t)\rangle$, respectively, 
in the region of Lyapunov indices 
indicating the stepwise structure of the Lyapunov spectra. 
   It should be emphasized that wavelike structure 
(the transverse Lyapunov modes) in the $y$-components 
of the spatial part and the momentum part of 
the Lyapunov vectors appears in the two-point steps of the Lyapunov 
spectra reported by \cite{Tan02c,Pos00}, 
and Fig. \ref{lyaveccomp} suggests that we may get a 
rather clear wave-like structure in the spatial 
part of the Lyapunov vectors than in its momentum part. 
   We can conclude a similar result for the four-point 
steps of the Lyapunov spectra in which time-depending  
wavelike-structure in the $y$-components 
of the spatial part of the Lyapunov vectors 
and in $x$-components of the spatial part of the 
Lyapunov vectors \cite{Tan02c,For02}.
   It may be noted that the time-averaged amplitudes 
$\langle\eta_{px}^{(n)}(t)\rangle$ and  
$\langle\eta_{py}^{(n)}(t)\rangle$, $n=2N-2$, $2N-1$ and 
$2N$ corresponding to zero-Lyapunov exponents are almost zero 
at any density (See the small panels in the top right of 
Figs. \ref{lyaveccomp}(a), (b), (c) and (d).).


\section{Density Dependence of the Largest Lyapunov Exponent and Related  Quantities}
\label{densitydependenceLmax}

   As we have already shown in Sect. \ref{Locza-Density}, 
the linear dependence [$\mathcal{LD}$] of localization widths 
as a function of Lyapunov index
appears in cases of low density, 
but it was not so clear at what density the linear dependence 
[$\mathcal{LD}$] of the Lyapunov localization starts to appear 
as the particle density $\rho$ deceases from 1. 
   In this section we discuss this problem 
by calculating density dependences 
of the largest Lyapunov exponent, 
the angle and amplitudes of Lyapunov vector components, 
and the localization width, which   
correspond to the largest Lyapunov exponent. 
   Especially, we show that the particle density region 
in which the linear dependence [$\mathcal{LD}$] appears is 
almost the same as the density region in which the largest 
Lyapunov exponent begins to satisfies the Krylov relation. 


\subsection{Largest Lyapunov Exponent}

   Fig. \ref{lyamaxdensi} is the largest Lyapunov exponent 
$\lambda^{(1)}$ as a function of the density $\rho$ 
 in the quasi-one-dimensional system. 
   In this figure we showed the numerical error 
$\epsilon\equiv \lambda^{(1)}+\lambda^{(4N)}$ 
(which must be zero in the Hamiltonian system) 
at each data point as an error bar 
of the length $2|\epsilon|$, showing that the values of 
the largest Lyapunov exponents at very low densities 
are less accurate than those at higher densities. 
   The grey region in this figure is the density region 
in which the linear dependence 
[$\mathcal{LD}$] of the localization widths appears. 

\begin{figure}[!htb]
\vspfigA
\caption{
      Largest Lyapunov exponent $\lambda^{(1)}$ 
   as a function of particle density $\rho$. 
      The grey region is the density region in which 
   the linear dependence [$\mathcal{LD}$] of the localization 
   widths as a function of Lyapunov index appears. 
      Numerical data is fitted by the Krylov relation 
   $\lambda^{(1)} \sim \alpha \rho \ln (\beta \rho)$ 
   (solid line) and the power function 
   $y=\alpha' \rho^{\beta'}$ (broken line) with 
   fitting parameters $\alpha$, $\beta$, 
   $\alpha'$ and $\beta'$. 
      The error bar in each data point is given by the 
   absolute value $2|\lambda^{(1)}+\lambda^{(4N)}|$ of twice 
   of the sum of the largest and smallest Lyapunov exponent.}
\vspfigB
\label{lyamaxdensi}\end{figure}  

   It is known that in the low density limit the largest Lyapunov 
exponent $\lambda^{(1)}$ should have the form 

\begin{eqnarray}
   \lambda^{(1)} \sim \alpha \rho \ln (\beta \rho) 
\label{Krylov}\end{eqnarray}

\noindent with parameters $\alpha$ and $\beta$. 
   Eq. (\ref{Krylov}) is called the Krylov relation \cite{Kry79} 
and has already been demonstrated numerically in a fully 
two-dimensional system 
consisting of many-hard-disks \cite{Del96,Del97b}, 
apart from a factor. 
   In Fig. \ref{lyamaxdensi} we fitted the numerical data 
to the Krylov relation (\ref{Krylov}) with parameter values 
$\alpha=-1.66875$ and $\beta=1.28252$ (the solid line), 
which gives a good fit for the density dependence of 
the largest Lyapunov exponent. 
   It is important to note that the density region 
that satisfies the Krylov relation almost coincides 
with the density region in which the linear dependence 
[$\mathcal{LD}$] of the localization widths appears. 
   To test the Krylov relation (\ref{Krylov}) we 
also tried to fit the numerical data by a power function 
$y=\alpha' \rho^{\beta'}$ (broken line) with fitting parameters  
$\alpha'$ and $\beta'$ in Fig. \ref{lyamaxdensi}. 
    We used the parameter values 
$\alpha'=2.55126$ and $\beta'=0.808158$, 
and it is shown in Fig. \ref{lyamaxdensi} 
that the Krylov relation (\ref{Krylov}) gives a better 
fit than this power function in the grey region. 

   Now we check the values of the fitting parameters 
$\alpha$ and $\beta$ used to fit the graph of 
Fig. \ref{lyamaxdensi} 
by following the rough derivation of the Krylov 
relation in Ref. \cite{Gas98}. 
   (Note that more exact derivations are known 
for some specific systems, such as  the
Lorentz gas, etc. 
   See, for example, Refs. \cite{Bei95,Chi97}.)
   First we note that after $n_{t}$ particle 
collisions the amplitude of a Lyapunov vector is 
stretched by a factor $(l_{f}/R)^{n_{t}}$ approximately 
with $l_{f}$ the mean free path. 
   Introducing the collision rate $\nu\equiv n_{t}/t$ to 
connect the mean free path $l_{f}$ with the time $t$ 
we estimate the largest Lyapunov exponent as 

\begin{eqnarray}
   \lambda^{(1)} \sim \lim_{t\rightarrow+\infty} 
   \frac{1}{t} \ln \left(\frac{l_{f}}{R}\right)^{n_{t}} 
   = \nu \ln \frac{l_{f}}{R}. 
\label{Krylov2}\end{eqnarray}

\noindent 
   We approximate the collision rate $\nu$ 
by a thermal velocity $u_{th}\equiv\sqrt{(2/M)(K/N)}$ 
as $\nu\sim u_{th}/l_{f}$ because of $u_{th} t\sim 
n_{t} l_{f}$, and assume that the mean free path 
$l_{f}$ is inversely proportional to the density $\rho$, 
so we obtain 
   
\begin{eqnarray}
   \lambda^{(1)} \sim - u_{th}\gamma \rho
   \ln (R\gamma \rho). 
\label{Krylov3}\end{eqnarray}

\noindent using a constant $\gamma$ of 
$l_{f}\sim (\gamma \rho)^{-1}$. 
   Comparing Eq. (\ref{Krylov3}) with Eq. (\ref{Krylov}) 
we obtain the relation 

\begin{eqnarray}
   \frac{\alpha}{\beta} \sim -\frac{u_{th}}{R}, 
\label{Krylov4}\end{eqnarray}

\noindent which is independent of the value of 
the constant $\gamma$. 
   Using the parameter values $u_{th}=\sqrt{2}$ and $R=1$ 
in our numerical calculations, we obtain 
$\alpha/\beta\sim-\sqrt{2}=-1.4142\cdots$, 
which is consistent of the values of the fitting parameters 
$\alpha$ and $\beta$ in Fig. \ref{lyamaxdensi}: 
$\alpha/\beta=-1.3011\cdots$.

\begin{figure}[!htb]
\vspfigA
\caption{
      Mean free time $\tau_{f}$ as a function of 
   density $\rho$.
      The dotted line is a plot of the fitting 
   function $y=1/(u_{th}\gamma x)$ with the thermal velocity 
   $u_{th}$ and the 
   fitting parameter $\gamma$. 
      The grey region is the density region in which 
   the linear dependence [$\mathcal{LD}$] of the  
   localization widths as a function of Lyapunov index 
   appears.   
   }
\vspfigB
\label{meanfreetime}\end{figure}  

   Next we proceed to estimate the factor 
$\gamma = (l_{f}\rho)^{-1}$. 
   Fig. \ref{meanfreetime} is the graph of the mean free time 
$\tau_{f}$ as a function of the density $\rho$. 
   Noting the relation $l_{f}\sim u_{th}\tau_{th}$, 
in this figure we fitted the numerical data by the function 
$y=1/(u_{th}\gamma x)$ with the value $\gamma =13.341$ 
for the fitting parameter $\gamma$. 
   (It may be emphasized that this function gives a 
nice fit in the density region in which the 
linear dependence [$\mathcal{LD}$] of the Lyapunov 
localization appears.)
   However this value $\gamma =13.341$, 
which is $\beta$ in Eq.  (\ref{Krylov}) in the case of $R=1$, 
is about 10 times as large as the fitting value of $\beta$ 
used to fit the graph of Fig. \ref{lyamaxdensi}. 
   It should be noted that because of the quasi-one-dimensional 
property of the system, 
a well known relation $l_{f}\sim 1/(\sqrt{2}\rho)$ 
\cite{Sea75} 
for fully two-dimensional hard-disk systems 
with a Maxellian velocity distribution cannot give 
a correct relation in the quasi-one-dimensional system, 
although this gives a parameter value 
$\beta$ which is rather close to the value used in 
Fig. \ref{lyamaxdensi}. 
   It may be noted that in order to derive the expression 
(\ref{Krylov3}) for the largest Lyapunov exponent we 
neglected some characteristics of the quasi-one-dimensional 
systems, such as the fact that in the quasi-one-dimensional system 
with a large value of the parameter $d$ particles collide mainly 
head on and it is rather rare for particles 
to have grazing collisions with other particles. 
   We should take these points into account to get 
more precise expressions for the parameters $\alpha$ and 
$\beta$ in Eq. (\ref{Krylov}) than in Eq. (\ref{Krylov3}).


\subsection{Angle and Amplitudes of Lyapunov Vector Components} 

   As the next example 
we consider the angle $\theta^{(1)}$ defined 
by Eq. (\ref{angleLyapuvecto}) between the spatial 
and momentum parts of the Lyapunov vector 
corresponding to the largest Lyapunov exponent. 
   Figure \ref{anglelyamax} is the graph of the time-average 
of the angle $\theta^{(1)}/\pi$ divided by $\pi$ 
as a function of particle density $\rho$. 
   The grey region in this figure  is the density region 
in which the linear dependence [$\mathcal{LD}$] of the  
localization widths appears. 
   The graph has a local maximum point about $\rho\approx 0.2$, 
which is close to the value of particle density in which 
the linear dependence [$\mathcal{LD}$] of the localization widths 
starts to appear as the density decreases from $1$. 
   As we have already discussed in Sects. \ref{LocalizationWidth} 
and \ref{Locza-Density}, the existence of the linear dependence 
[$\mathcal{LD}$] of the localization widths can be checked 
by a linear dependence of the localization widths as a function of 
Lyapunov index and a rectangular shape of the amplitude of the 
Lyapunov vector, but it is rather hard to use this criterion 
to distinguish the density region of the linear dependence 
[$\mathcal{LD}$], because it is not easy to recognize these behaviors 
in an intermediate region between the density region of 
the exponential dependence [$\mathcal{ED}$] only 
and the density region of  
both the dependences [$\mathcal{ED}$] and [$\mathcal{LD}$].  
   In this sense the graph of the angle $\theta^{(1)}$ 
as a function of particle density $\rho$ may give a more distinct 
criterion to distinguish the density region in which the 
linear dependence [$\mathcal{LD}$] appears. 
   It is noted that the angle $\theta^{(1)}$ does not seems to go 
to zero in the low density limit as shown in the inset of Fig. 
\ref{anglelyamax}.

\begin{figure}[!htb]
\vspfigA
\caption{
      Density dependence of the time-average 
   $\langle\theta^{(1)}\rangle/\pi$ 
   of the angle divided by $\pi$ 
   between the spatial part and the momentum part 
   of the Lyapunov vector corresponding to the largest Lyapunov 
   exponent as a function of the particle density $\rho$. 
      The grey region is the density region in which 
   the linear dependence [$\mathcal{LD}$] of the  
   localization widths as a function of Lyapunov index appears. 
      The inset: The same graph but including the angle 
   $\theta^{(1)}$ 
   at lower density.}
\vspfigB
\label{anglelyamax}\end{figure}  

   Next we consider the density dependence of the time-average 
of normalized amplitude of the $x$-component and the $y$-component 
in the spatial part and the momentum part of the 
Lyapunov vector corresponding to the largest Lyapunov exponent, 
which are   
defined by Eqs. (\ref{lyaamp2}) and (\ref{lyaamp3}) in $n=1$. 
   Fig. \ref{lyaveccomplmax} is the graphs of 
$\langle\eta_{qx}^{(1)}(t)\rangle$ (circles), 
$\langle\eta_{qy}^{(1)}(t)\rangle$ (triangles), 
$\langle\eta_{px}^{(1)}(t)\rangle$ (crosses) and 
$\langle\eta_{py}^{(1)}(t)\rangle$ (diagonal crosses) 
as functions of the particle density $\rho$. 
   Again the density region in which the linear dependence 
[$\mathcal{LD}$] of the localization widths appears is indicated 
as a grey region. 
   This may not be a good criterion to distinguish the 
density region of the linear dependence [$\mathcal{LD}$], 
but it is clear that in the density region of the 
linear dependence [$\mathcal{LD}$] 
the spatial parts of the Lyapunov vector is 
dominant and a gap appears between its $x$-component 
$\langle\eta_{qx}^{(1)}(t)\rangle$ and the $y$-component 
$\langle\eta_{qy}^{(1)}(t)\rangle$. 
   On the other hand, in a high 
density region in which the linear dependence 
[$\mathcal{LD}$] does not appear, the momentum parts 
$\langle\eta_{px}^{(1)}(t)\rangle$ and 
$\langle\eta_{py}^{(1)}(t)\rangle$ of the Lyapunov vector 
are dominant. 
   It should be noted that in the low density 
in which the linear dependence [$\mathcal{LD}$] appears, 
the transverse components $\langle\eta_{qy}^{(1)}(t)\rangle$ and 
$\langle\eta_{py}^{(1)}(t)\rangle$ are always larger than 
the longitudinal components $\langle\eta_{qx}^{(1)}(t)\rangle$ 
and $\langle\eta_{px}^{(1)}(t)\rangle$, respectively.

\begin{figure}[!htb]
\vspfigA
\caption{
      Time-averages of the normalized amplitudes of 
   the $x$-component of the spatial part 
   ($\langle\eta_{qx}^{(1)}(t)\rangle$, circles), 
   the $y$-component of the spatial part 
   ($\langle\eta_{qy}^{(1)}(t)\rangle$, triangles), 
   the $x$-component of the momentum part 
   ($\langle\eta_{px}^{(1)}(t)\rangle$, crosses), 
   and the $y$-component of the momentum part 
   ($\langle\eta_{py}^{(1)}(t)\rangle$, diagonal crosses) 
   of the Lyapunov vector corresponding 
   to the largest Lyapunov exponent as functions of 
   the particle density $\rho$.
      The grey region is the density region in which 
   the linear dependence [$\mathcal{LD}$] of the  
   localization widths as a function of Lyapunov index appears. }
\vspfigB
\label{lyaveccomplmax}\end{figure}  


\subsection{Localization Width} 
\label{localizationwidthcorresponding}

   As the last quantity in this section, 
we consider the density dependence of the Lyapunov 
localization width corresponding to the largest Lyapunov exponent. 
   Figure \ref{loczawidthlyamax} is the graph of 
the Lyapunov localization width $\mathcal{W}^{(1)}/N$ 
normalized by the particle number $N$ as a function of density $\rho$. 
   This figure shows that in the low density limit the normalized 
localization width $\mathcal{W}^{(1)}/N$ goes to the value 
$2/N (=0.04)$ (solid line in Fig. \ref{loczawidthlyamax}), 
which is discussed as the minimum localization width 
$\mathcal{W}_{min}/N$ in Sect. \ref{LocalizationWidth}. 
   In Fig. \ref{loczawidthlyamax} we indicate the density region 
of the linear dependence [$\mathcal{LD}$] as the grey region, but 
it is not so clear from this figure how to distinguish the density region 
in which the linear dependence [$\mathcal{LD}$] of the localization 
widths appears. 

\begin{figure}[!htb]
\vspfigA
\caption{
      Normalized localization width $\mathcal{W}^{(1)}/N$ 
   corresponding to the largest Lyapunov 
   exponent as a function of density $\rho$. 
      The solid line is the minimum value of the 
   normalized localization width: 
   $\mathcal{W}_{min}/N =0.04$. 
      The grey region is the density region in which 
   the linear dependence [$\mathcal{LD}$] of the  
   localization widths as a function of Lyapunov index appears.}
\vspfigB
\label{loczawidthlyamax}\end{figure}  


\section{Conclusion and Remarks}
\label{ConclusionRemarks}

   In this paper we have discussed localized behaviors 
 of Lyapunov vectors (the Lyapunov localization)
for quasi-one-dimensional systems 
consisting of many hard disks 
(with periodic boundary conditions except 
in Appendix \ref{AppendixHardwall}).  
   The quasi-one-dimensional system was introduced as 
a particle system whose shape is a very narrow rectangle 
that does not allow the exchange of particle positions. 
   We compared some methods to characterize 
the localized behavior of the Lyapunov vectors, and 
as one of such methods in this paper 
we used a quantity 
called the localization width, whose logarithm is given 
by an entropy for the amplitude distribution of 
the Lyapunov vector components of each particle. 
   The localization width indicates the number of particles 
contributing to the localized part of Lyapunov vector components. 
   It could not only be used as an indicator to measure 
the magnitude of the localized behavior of the Lyapunov 
vectors, but also be used to distinguish different  
delocalized properties of the Lyapunov vectors 
such as the delocalization associated with a random 
distribution of particle component amplitudes, a delocalization associated 
with a uniform distribution and a delocalization 
associated with a wave-like structure 
(corresponding to stepwise structure 
of the Lyapunov spectra ) in the Lyapunov vector 
(Figs. \ref{loczwidndep} and \ref{lypuaNdepen}). 
   The localized region of the Lyapunov vectors 
is related to the points of colliding particles 
(Fig. \ref{loczaamplicolli}), and 
this leads to the lower bound $2$ of the 
localization width 
(Figs. \ref{lyapuwidthlow} and \ref{loczawidthlyamax}). 
   Using the localization width we showed that there are two 
kind of the Lyapunov localizations in many-hard-disk systems. 
   The first type of the Lyapunov localization 
is characterized by an exponential dependence [$\mathcal{ED}$] 
of the localization widths as a function of Lyapunov index 
(Fig. \ref{lyapuwidthhigh}), 
and by its long tale of localized 
Lyapunov vectors (Fig. \ref{loczat}(a)). 
   This type of the Lyapunov localization is observed 
at any particle density. 
   The second type of the Lyapunov localization 
is characterized by the linear dependence [$\mathcal{LD}$] of the 
localization widths as a function of Lyapunov index 
(Fig. \ref{lyapuwidthlow}), 
and by the sharp rectangular shape of the localized 
Lyapunov vectors  (Figs. \ref{loczat}(b) and \ref{loczat2}). 
   This type of Lyapunov localization appears 
only in low density cases and in Lyapunov indices 
corresponding to the large Lyapunov exponents in absolute values. 
   We showed that in the density region of 
the linear dependence [$\mathcal{LD}$] of the localization widths 
the Lyapunov spectra is bent and separated into 
two parts (except for the stepwise region of the Lyapunov 
spectra): one corresponding to the exponential dependence 
[$\mathcal{ED}$] and take very small values corresponding to 
the largest Lyapunov exponent, and the other corresponding 
to the linear dependence [$\mathcal{LD}$] and shows a rapid 
decreasing dependence of  
Lyapunov index (Fig. \ref{lyaspe}). 
   It was also shown that differences between 
the exponential dependence [$\mathcal{ED}$] and the linear 
dependence  
[$\mathcal{LD}$] appear in the angle $\theta^{(n)}$ 
between the spatial 
and momentum parts of the Lyapunov vectors 
(Fig. \ref{anglyavec})
and in the amplitudes of the $x$-component and 
$y$-component of the spatial part (Fig. \ref{lyaveccomp}(d)). 
   (Here we took the $y$-direction as the narrow direction of 
the rectangle and the $x$-direction as the longer orthogonal 
direction.) 
   The density region, in which the linear dependence [$\mathcal{LD}$] 
of localization widths appears, almost exactly coincides 
with the density region in which the density dependence 
of the largest Lyapunov exponent $\lambda^{(1)}$ satisfies the 
Krylov relation (Fig. \ref{lyamaxdensi}). 
   We also indicated that at the boundary of 
the density region of the exponential dependence [$\mathcal{ED}$] only 
and the density region of both the linear dependence 
[$\mathcal{LD}$] and the exponential dependence [$\mathcal{ED}$],  
the angle $\theta^{(1)}$ corresponding to the 
largest Lyapunov exponent shows a local maximum 
as a function of particle density (Fig. \ref{anglelyamax}).  

   In this paper we observed differences in the amplitudes 
of the $x$ and $y$-components of the 
Lyapunov vectors 
(Figs. \ref{lyaveccomp} and \ref{lyaveccomplmax}). 
   These differences come from the difference in the roles 
of the directions in the quasi-one-dimensional systems. 
   We also observed differences in the amplitudes 
of the spatial and momentum parts of the Lyapunov vectors.  
   In the region where the 
Lyapunov spectra are changing smoothly the amplitude of the 
spatial part of the Lyapunov vectors is 
larger (smaller) than of the momentum part 
in low (high) density cases 
(Figs. \ref{lyaveccomp}, \ref{lyaveccomplmax} 
and \ref{lyaveccompsqu}). 
   The spatial and momentum parts of the Lyapunov vectors are 
in almost the same direction at high density, 
whereas they are rather close to orthogonal 
in the low density case, especially in the region of 
the exponential dependence [$\mathcal{ED}$] of the 
localization widths as a function of Lyapunov index 
(Figs. \ref{anglyavec} and \ref{anglyasqu}). 
   These behaviors are found not only in the 
quasi-one-dimensional systems but also in the square system 
(Fig. \ref{lyaveccompsqu}). 
   Concerning the stepwise region of the Lyapunov spectra, 
as shown in Fig. \ref{lyaveccomp}, 
the amplitudes of the $y$-component, 
(transverse component), of the spatial 
and momentum parts of the Lyapunov vectors are larger 
than the corresponding $x$-components, 
namely (their longitudinal components) 
in the two-point steps of the Lyapunov spectra, whereas 
they are opposite in four-point steps of the Lyapunov spectra 
(except in very low density cases like 
in Fig. \ref{lyaveccomp}(d)).

   The particle number dependence of the Lyapunov localization 
width normalized by the particle number $N$ was investigated 
a little in this paper. 
   Figure \ref{loczwidndep} suggests that the 
Lyapunov localization width normalized by the particle number 
decreases as a function of the particle number itself, but 
we need a further calculation to show the existence of 
its thermodynamical limit ($N\rightarrow+\infty$), 
as well as its low density limit ($\rho\rightarrow 0$). 

   We suggested that the bending of the Lyapunov spectra 
in the low density cases accompanying the linear dependence 
[$\mathcal{LD}$] of the localization widths may come 
from a time-scale separation of the dynamics. 
   As one of the possible to explanations of this point 
we demonstrate a strong asymmetry of the amplitudes 
of the spatial and momentum parts of the 
Lyapunov vectors at low density. 
   However this may not give an enough explanation as to 
where the bending point of the Lyapunov spectra is, 
and one may also indicate that 
the Lyapunov spectrum is related to growing (or reducing) 
speeds of the 
amplitudes of the Lyapunov vectors, not the amplitudes 
of the Lyapunov vectors themselves. 
   This remains an open question. 

   Results in this paper suggest that the localized behavior of 
the Lyapunov vectors comes from the short range property 
of particle interaction (Fig. \ref{loczaamplicolli}), 
noting that the hard-core interaction 
of the systems used in this paper is the shortest range interactions 
possible.  
   Therefore it should be interesting to compare our results 
with the Lyapunov localization of systems with long 
range particle interactions. 
   Moreover, if the Lyapunov localization is connected 
to the existence of the thermodynamical limit of the 
largest Lyapunov exponent \cite{Mil02}, 
then it should be interesting 
to investigate the Lyapunov localization in the 
Fermi-Pasta-Ulam model in which a numerical calculation 
suggests a logarithmic divergence of the largest Lyapunov 
exponent as a function of particle number $N$ \cite{Sea97}. 
   These remain as future problems 
on the subject of Lyapunov localizations.


\section*{Acknowledgements}

   One of the author (T.T) wish to thank to T. Harayama and 
H. Schomerus for valuable information about localization problems. 
   We are grateful for financial support for this work 
from the Australian Research Council. 


\appendix

\section{Inequality for the Localization Width} 
\label{appen0}

   In this appendix we derive the inequality (\ref{LoczatWidthPropa})  
for the localization width of the Lyapunov vector. 
   
   First we note the inequality $0 \leq S^{(n)}$ 
which comes from Eqs. (\ref{lyaampC}) and (\ref{LoczatEntro}). 
   This leads to the inequality $1 \leq \mathcal{W}^{(n)}$.
   
   Second, we note 

\begin{eqnarray}
   &\ln N& + \sum_{j=1}^{N} \gamma_{j}^{(n)}(t) 
      \ln \gamma_{j}^{(n)}(t) \nonumber \\
   && = N^{-1}\sum_{j=1}^{N} 
      \left[\gamma_{j}^{(n)}(t) N\right] 
      \ln \left[\gamma_{j}^{(n)}(t) N\right] 
      \nonumber \\
   && = N^{-1}\sum_{j=1}^{N} 
      \left\{\left[\gamma_{j}^{(n)}(t) N\right] 
      \ln \left[\gamma_{j}^{(n)}(t) N\right] 
      \right. \nonumber \\
   &&\hspace{2cm} \left. - \left[\gamma_{j}^{(n)}(t) N \right] 
      +1 \right\}
      \nonumber \\
   && = N^{-1}\sum_{j=1}^{N} \mathcal{F}\left( 
      \gamma_{j}^{(n)}(t) N\right) 
\label{derivineq1}\end{eqnarray}

\noindent where we used Eq. (\ref{lyaampB}), and the function 
$\mathcal{F}(x)$ of $x$ is defined by 

\begin{eqnarray}
   \mathcal{F}(x) \equiv x \ln x -x +1 . 
\label{derivineq2}\end{eqnarray}

\noindent It is easy to show that the function 
$\mathcal{F}(x)$ satisfies the inequality  

\begin{eqnarray}
   \mathcal{F}(x) \geq 0, \hspace{1cm} \mbox{in} \;\;\; x\geq 0 .
\label{derivineq3}\end{eqnarray}

\noindent Eqs. (\ref{LoczatEntro}) and 
(\ref{derivineq1}), and the inequality (\ref{derivineq3}) 
lead to

\begin{eqnarray}
   \ln N - S^{(n)} = 
      N^{-1}\sum_{j=1}^{N} \left\langle \mathcal{F}\left( 
      \gamma_{j}^{(n)}(t) N\right) \right\rangle \geq 0 ,
\label{derivineq4}\end{eqnarray}

\noindent noting the inequality $\gamma_{j}^{(n)}(t) N\geq 0$. 
   Therefore we obtain the inequality 
$\mathcal{W}^{(n)} \leq N$ using Eq. (\ref{LoczatWidth}). 
   From the two inequalities $1 \leq \mathcal{W}^{(n)}$
and $ \mathcal{W}^{(n)} \leq N$ 
we derive the inequality (\ref{LoczatWidthPropa}) 
for the localization width. 

   We can also show that the function 
$\mathcal{F}(x)$ defined by Eq. (\ref{derivineq2}) satisfies 
$\mathcal{F}(x) = 0$ only when $x=1$. 
   Using this point and Eqs. (\ref{LoczatWidth}) 
and (\ref{derivineq4}) we get 
the fact that the equality $\mathcal{W}^{(n)} = N$ 
for the localization width is satisfied only when all the quantities 
$\gamma_{j}^{(n)}(t)$, $j=1,2,\cdots N$ are equal, namely when  
$\gamma_{j}^{(1)}(t) = \gamma_{j}^{(2)}(t) 
= \cdots=\gamma_{j}^{(N)}(t) = 1/N$.


\section{Comparison with Square Cases} 
\label{appenA}

   In this appendix we discuss differences between 
quasi-one-dimensional systems 
and square systems for the Lyapunov localization 
and its related phenomena. 
   For the quasi-one-dimensional system in this paper we chose  
the system lengths as 

\begin{eqnarray}
   L_{y} &=& L_{y}' \equiv 2R(1+10^{-6}), \\
   L_{x} &=& L_{x}' \equiv NL_{y}(1+d) \\
   && \mbox{\hspace{-1.2cm} [in the quasi-one-dimensional cases]}
   \nonumber
\end{eqnarray}

\noindent with a parameter $d$ to change the 
particle density. 
   For meaningful comparisons between the square and  
quasi-one-dimensional cases we use system lengths 
for the square systems so that both cases give the same area, 
for the same value of the parameter $d$, namely 

\begin{eqnarray}
   L_{y} &=& L_{x} = \sqrt{L_{x}'L_{y}'} 
   \nonumber \\
   &=& \sqrt{N(1+d)[2R(1+10^{-6})]^2}\\
   && \mbox{\hspace{-1.2cm} [in the square cases].}
   \nonumber
\end{eqnarray}

\noindent
   Except for these lengths $L_{x}$ and $L_{y}$ 
we use the same parameter
values as given in the text, such as  $R=1$, $M=1$ and $E=N$.  
   In this appendix we consider systems of $50$ particles ($N=50$).  
   In Appendixes \ref{AppendixLyapunovSpectra}, 
\ref{AppendixLocalizationWidths}, 
\ref{AppendixAngleLyapunovVector} and 
\ref{AppendixAmplitudesofXcomponents} 
we consider the case of $d=10^{5}$ 
in which the linear dependence [$\mathcal{LD}$] 
of localization widths with respect to 
Lyapunov index appears. 
   In Appendix \ref{AppendixHardwall} we discuss boundary effects 
in the localization width in the case of $d=0.5$.


\subsection{Lyapunov Spectra}
\label{AppendixLyapunovSpectra}

   The first example is the Lyapunov spectra of the 
quasi-one-dimensional case and the square case. 
   Fig. \ref{lyapusqu} is the positive branch of 
the Lyapunov spectra normalized 
by the largest Lyapunov exponent $\lambda^{(1)}$ as functions 
of the normalized Lyapunov index $n/(2N)$ in the quasi-one-dimensional 
system (circles) and the square system 
(triangles) in the case of $d=10^{5}$. 
   Here the values of the largest Lyapunov exponents are 
given by 
$\lambda^{(1)} \approx 0.000157$ in the quasi-one-dimensional system, 
and $\lambda^{(1)} \approx 0.000552$ in the square system.  
   As shown in Fig. \ref{lyapusqu}, 
a sharp bending of the Lyapunov spectra occurs 
in both cases, but  
it occurs in a slightly smaller Lyapunov index 
in the square case than in the quasi-one-dimensional case.  
   The Lyapunov spectrum in the quasi-one-dimensional 
case is also shown in Fig. \ref{lyaspe}.

\begin{figure}[!htb]
\vspfigA
\caption{
      Lyapunov spectra normalized by the largest Lyapunov exponent 
   $\lambda^{(1)}$ as functions of the normalized Lyapunov index 
   $n/(2N)$ in the case of $d=10^{5}$
   in the quasi-one-dimensional 
   system (circles) and the square system  
   (triangles). 
      Inset: Enlarged graphs of the normalized 
   Lyapunov spectra in a region of small positive 
   Lyapunov exponents.}
\vspfigB
\label{lyapusqu}\end{figure}  

   There is a difference between the quasi-one-dimensional case and 
the square cases in the region of small positive Lyapunov exponents.  
   In the inset of Fig. \ref{lyapusqu}, we can recognize two 
stepwise structures of the Lyapunov spectrum consisting 
of six Lyapunov exponents in the quasi-one-dimensional system. 
   It should be noted that each of these steplike structures 
consists of one four-point step and one two-point step 
although these two steps are too close to be distinguished 
in Fig. \ref{lyapusqu}. 
   (This is also evidenced by investigating wavelike structures 
of the Lyapunov vectors, the so called Lyapunov modes 
\cite{Tan02c}.)
   On the other hand we cannot recognize such a stepwise 
structure in the square case, because $50$ particles 
in the square system with 
the density $\rho=0.00000785\cdots$ 
are too small to show stepwise structure of the 
Lyapunov spectrum. 
   (It is known that rectangular systems including the 
quasi-one-dimensional systems show a longer 
stepwise region of Lyapunov spectra than in the 
square system with the same area \cite{Pos00}.) 
   In the square system the gap between the smallest 
non-zero positive Lyapunov exponent $\lambda_{2N-4}$ 
and the zero Lyapunov exponents is larger than in the 
quasi-one-dimensional system.


\subsection{Localization Widths}
\label{AppendixLocalizationWidths}

   Next we consider the localization width 
in the quasi-one-dimensional case and the square case. 
   Fig. \ref{loczwidsqu} is the localization widths 
$\mathcal{W}^{(n)}/N$, $n=1,2,\cdots,2N$ normalized by 
the particle number $N$  
as functions of the normalized Lyapunov index $n/(2N)$ 
in the quasi-one-dimensional case 
(circles) and the square case (triangles)
in the case of $d=10^{5}$. 
   In the region of Lyapunov index corresponding to the Lyapunov 
spectra changing smoothly, qualitative behavior of the 
localization width are rather similar, and consist of 
the linear dependence [$\mathcal{LD}$] in small Lyapunov indices 
and the exponential dependence [$\mathcal{ED}$] in other 
indices,  although 
the region of the linear dependence [$\mathcal{LD}$] in the square case 
is slightly smaller than in the quasi-one-dimensional case. 
   In Fig. \ref{loczwidsqu} we give only fits of 
exponential functions $y=\tilde{\alpha} + 
\tilde{\beta}\exp\{\tilde{\gamma} x\}$ 
for the exponential dependence [$\mathcal{ED}$] by dotted lines.  
   (The parameter values to fit data are 
$(\tilde{\alpha},\tilde{\beta},\tilde{\gamma}) 
= (0.560814, -2.73348, -7.50544)$ 
for the quasi-one-dimensional case, and 
$(\tilde{\alpha},\tilde{\beta},\tilde{\gamma}) 
= (0.597656, -3.35386, -10.9674)$ for the square case.) 
   The localization widths are larger than 
their minimum value $\mathcal{W}^{(n)} = \mathcal{W}_{min}=2$ 
and are smaller than 
its random component case $\mathcal{W}^{(n)} = \mathcal{W}_{ran} 
\approx 0.651N$ except in the stepwise region of the Lyapunov 
spectra, and the smallest localization width 
$\mathcal{W}^{(1)}$ are close to $\mathcal{W}_{min}$ in both the cases. 

\begin{figure}[!htb]
\vspfigA
\caption{
      Normalized localization width 
   $\mathcal{W}^{(n)}/N$ 
   as functions of the normalized Lyapunov index 
   $n/(2N)$ in the case of $d=10^{5}$
   in the quasi-one-dimensional 
   case (circles) and the square case (triangles). 
      The dotted lines are fits by exponential functions 
   (the dependence [$\mathcal{ED}$]). 
      The broken line interrupted by dots, 
   the solid line and the broken line 
   correspond to  
   $\mathcal{W}^{(n)} = \mathcal{W}_{wav} (\approx 0.736N)$, 
   $\mathcal{W}_{ran} (\approx 0.651N)$ 
   and $\mathcal{W}_{min} (=2)$, 
   respectively. 
   }
\vspfigB
\label{loczwidsqu}\end{figure}  

   As in the Lyapunov spectra, the localization width shows 
a different behavior with the Lyapunov index corresponding to 
the stepwise structure of the Lyapunov spectra. 
   As we have already mentioned, in the square case of $N=50$ 
there is no stepwise region of the Lyapunov spectrum, 
so there is not a corresponding structure in the localization 
width. 
   We can see that in the square case 
the localization widths $\mathcal{W}^{(n)}$ 
decrease as a function of the Lyapunov index $n$ in the region 
of large Lyapunov indices corresponding to small positive 
Lyapunov exponents. 
   On the other hand we can see the effect of the stepwise structure 
of the Lyapunov spectrum on the localization width 
of the quasi-one-dimensional case. 
   Especially a pair of two dots on the line 
$\mathcal{W}^{(n)}=\mathcal{W}_{wav}$ correspond to 
the two-point step of the Lyapunov spectrum shown in Fig. 
\ref{lyapusqu}, supporting that in the Lyapunov spectrum of 
Fig. \ref{lyapusqu} for the quasi-one-dimensional case 
the first (second) step of the Lyapunov spectrum is 
a four-point step (a two-point step) looking from the 
zero-Lyapunov exponents. 


\subsection{Angle Between the Spatial and Momentum Parts of the Lyapunov Vector} 
\label{AppendixAngleLyapunovVector}

   Fig. \ref{anglyasqu} is the time-averaged angles 
$\langle\theta^{(n)}\rangle/\pi$, $n=1,2,\cdots,2N$ 
divided by $\pi$ between the spatial and momentum parts 
of the Lyapunov vectors in the cases of the quasi-one-dimensional system 
(circles) and the square system (triangles) 
as functions of the normalized Lyapunov index $n/(2N)$ 
for $d=10^{5}$. 
   In both the cases, the angle $\theta^{(n)}$ is a rapidly 
increasing function of Lyapunov index $n$ in the linear 
dependence [$\mathcal{LD}$] region of the localization width, and 
almost $\pi/2$  in the exponential dependence [$\mathcal{ED}$] 
region. 
   
\begin{figure}[!htb]
\vspfigA
\caption{
      Time-average  $\langle\theta^{(n)}\rangle/\pi$ 
   of the angles divided by $\pi$ 
   between the spatial and momentum 
   parts of the Lyapunov vectors 
   in the quasi-one-dimensional 
   case (circles) and the square case (triangles) 
   as functions of the normalized Lyapunov index 
   $n/(2N)$ for $d=10^{5}$. 
      The solid line corresponds to 
   the value $\theta^{(n)}=\pi/2$ of the angle. 
      Inset: Enlarged graphs of normalized angles 
   $\langle\theta^{(n)}\rangle/\pi$ 
   in the region of Lyapunov index in which 
   the the exponential dependence [$\mathcal{ED}$] 
   of the localization widths (and the 
   stepwise structure of the Lyapunov spectrum) 
   appears.}
\vspfigB
\label{anglyasqu}\end{figure}  

   We can see the stepwise structure in the angle $\theta^{(n)}$ 
as a function of Lyapunov index $n$ in the quasi-one-dimensional 
case (See the inset of Fig. \ref{anglyasqu}.). 
   This structure corresponds to the stepwise structure of the 
Lyapunov spectrum shown in Fig. \ref{lyapusqu}.


\subsection{Amplitudes of the $x$- and $y$-Components of the Spatial and Momentum Parts of the Lyapunov Vectors} 
\label{AppendixAmplitudesofXcomponents}

   As the next quantities to investigate the difference 
between the quasi-one-dimensional case and the square case, 
we consider the amplitudes of the $x$- and $y$-components 
of the spatial and momentum parts 
of the Lyapunov vectors. 
   Fig. \ref{lyaveccompsqu} is the graphs of the time-averages 
of normalized amplitudes of 
the $x$-component of the spatial part 
 ($\langle\eta_{qx}^{(n)}(t)\rangle$, circles), 
the $y$-component of the spatial part 
($\langle\eta_{qy}^{(n)}(t)\rangle$, triangles), 
the $x$-component of the momentum part
($\langle\eta_{px}^{(n)}(t)\rangle$, crosses), 
and its $y$-component 
($\langle\eta_{py}^{(n)}(t)\rangle$, diagonal crosses) 
of the Lyapunov vectors 
as functions of normalized Lyapunov index $n/(2N)$ 
in the square case for $d=10^{5}$. 
   The corresponding graphs for 
the quasi-one-dimensional case are shown 
in Fig. \ref{lyaveccomp}(d). 
   In the square system there is no difference 
between the $x$- and $y$-directions, 
so there is no gap between the $x$-
and y-components in the spatial part 
and the momentum part of the Lyapunov vectors 
(except for those where the spatial part 
corresponds to the zero-Lyapunov exponents). 
   Although the momentum parts 
$\langle\eta_{px}^{(n)}(t)\rangle$ 
and $\langle\eta_{py}^{(n)}(t)\rangle$ 
(the spatial parts 
$\langle\eta_{qx}^{(n)}(t)\rangle$ and
$\langle\eta_{qy}^{(n)}(t)\rangle$) are rapidly decreasing 
(increasing) functions of Lyapunov index in the region 
of the linear dependence [$\mathcal{LD}$] of the localization 
widths 
(See the small figure in the top right 
of Fig. \ref{lyaveccompsqu} for 
the momentum parts of the Lyapunov vectors. 
Also note the normalization condition (\ref{lyaamp23nor}) 
to know the corresponding spatial parts.), 
the boundary of  
the two dependences [$\mathcal{LD}$] and [$\mathcal{ED}$] 
is not so clear in this figure. 

%
\begin{figure}[!htb]
\vspfigA
\caption{
      Time-average of the normalized amplitudes of 
   the $x$-component of the spatial part 
   ($\langle\eta_{qx}^{(n)}(t)\rangle$, circles), 
   the $y$-component of the spatial part 
   ($\langle\eta_{qy}^{(n)}(t)\rangle$, triangles), 
   the $x$-component of the momentum part
   ($\langle\eta_{px}^{(n)}(t)\rangle$, crosses), 
   and the $y$-component of the momentum part 
   ($\langle\eta_{py}^{(n)}(t)\rangle$, diagonal crosses) 
   of the Lyapunov vectors 
   as functions of normalized Lyapunov index $n/(2N)$ 
   in the square case for $d=10^{5}$. 
   Corresponding quasi-one-dimensional case is given 
   in Fig. \ref{lyaveccomp}(d). 
      The small panels in the top right 
   are the graphs of $\langle\eta_{px}^{(n)}(t)\rangle$ 
   and $\langle\eta_{py}^{(n)}(t)\rangle$ only. 
   }
\vspfigB
\label{lyaveccompsqu}\end{figure}  

   In Fig. \ref{lyaveccompsqu} we cannot recognize 
a stepwise structure of the quantities 
$\langle\eta_{qx}^{(n)}(t)\rangle$, 
$\langle\eta_{qy}^{(n)}(t)\rangle$, 
$\langle\eta_{px}^{(n)}(t)\rangle$ 
and $\langle\eta_{py}^{(n)}(t)\rangle$, 
because there is no stepwise structure of the 
Lyapunov spectrum in this system.
   It may be noted that the normalized $x$-component 
$\langle\eta_{px}^{(n)}(t)\rangle$ and 
the normalized $y$-component 
$\langle\eta_{py}^{(n)}(t)\rangle$ of the momentum part 
of the Lyapunov vectors corresponding to the zero-Lyapunov 
exponents $\lambda^{(2N-2)}$, $\lambda^{(2N-1)}$ and 
$\lambda^{(2N)}$ are almost zero, like in the 
quasi-one-dimensional case.


\subsection{Boundary Effects in the Localization Widths}
\label{AppendixHardwall}

   As the last example in this appendix we check 
boundary effects in Lyapunov localizations. 
   In this subsection we calculate the localization widths 
for square systems and quasi-one-dimensional systems 
with purely hard-wall boundary conditions or 
purely periodic boundary conditions. 
   To give reasonable comparisons between the 
the hard-wall boundary cases and the periodic boundary 
cases we choose the system lengths $L_{x}$ and $L_{y}$ so that 
the effective region for a particle to move is the same 
in both cases. 
   This means that if we choose the system size as 
$L_{x}=\tilde{L}_{x}$ and $L_{y}=\tilde{L}_{y}$ for the 
periodic boundary case then they must be 
$L_{x}=\tilde{L}_{x}+2R$ and $L_{y}=\tilde{L}_{y}+2R$ 
for the corresponding hard-wall boundary case. 
   Except for this point, we chose the same value 
of the other parameters for 
these two boundary cases.  
   In this subsection we use the value $d=0.5$ for the parameter $d$. 
   Note that  in this case the linear dependence [$\mathcal{LD}$] of the 
localization widths does not appear. 

\begin{figure}[!htb]
\vspfigA
\caption{
      Normalized localization widths $\mathcal{W}^{(n)}/N$ 
   as functions of the normalized Lyapunov index $n/(2N)$
   for $d=0.5$
   in a square system 
   with periodic boundary conditions (circles), 
   in a square system 
   with hard-wall boundary conditions (triangles), 
   in a quasi-one-dimensional system 
   with periodic boundary conditions (crosses), 
   in a quasi-one-dimensional system 
   with hard-wall boundary conditions (diagonal crosses). 
     The dotted lines are fits of the 
   numerical data by exponential functions. 
      The broken line interrupted by dots, 
   the solid line and the broken line 
   correspond to  
   $\mathcal{W}^{(n)} = \mathcal{W}_{wav} (\approx 0.736N)$, 
   $\mathcal{W}_{ran} (\approx 0.651N)$ 
   and $\mathcal{W}_{min} (=2)$, 
   respectively. 
    }
\vspfigB
\label{boundary}\end{figure}  

   Figure \ref{boundary} is the localization widths 
$\mathcal{W}^{(n)}/N$, $n=1,2,\cdots,2N$ normalized by 
the particle number $N$ in the case of 
the square system with the periodic boundary conditions (circles), 
the square system with the hard-wall boundary conditions (triangles), 
the quasi-one-dimensional system with the periodic 
boundary conditions (crosses), 
the quasi-one-dimensional system with the hard-wall 
boundary conditions (diagonal crosses). 
   Fig. \ref{boundary} shows that the localization widths 
in the hard-wall boundary conditions are smaller than 
in the periodic boundary conditions in both of 
the square cases and the quasi-one-dimensional cases. 
   However it should be noted that differences in the values 
of the localization widths are rather small in the region of 
small Lyapunov indices $n$ in which the Lyapunov vectors are 
strongly localized. 
   This is a natural result because the Lyapunov localization 
is a local behavior so that it should not depend strongly 
on global conditions such as the boundary conditions.    
   It may be noted that in very narrow systems like 
the quasi-one-dimensional system a strong boundary effect 
appears because particles can collide with walls 
much more often than in square cases, and this could be the reason 
why the differences in the values 
of the localization widths from the different boundary conditions 
are larger in the quasi-one-dimensional system than in the 
square system. 
   A part of the localization widths 
as a function of Lyapunov index are nicely fitted by the 
exponential function $y=\tilde{\alpha} + 
\tilde{\beta}\exp\{ \tilde{\gamma} x\}$ 
for the exponential dependence [$\mathcal{ED}$] with 
fitting parameters $\tilde{\alpha}$, 
$\tilde{\beta}$ and $\tilde{\gamma}$, 
shown as the dotted lines in Fig. \ref{boundary}. 
   Here, the values of fitting parameters are chosen as   
$(\tilde{\alpha},\tilde{\beta},\tilde{\gamma}) = 
(0.653629, -0.377495, -10.9742$ for 
the square case with the periodic boundary conditions,  
$(\tilde{\alpha},\tilde{\beta},\tilde{\gamma}) = 
(0.638696, -0.422529, -10.3554)$ for 
the square case with the hard-wall boundary conditions, 
$(\tilde{\alpha},\tilde{\beta},\tilde{\gamma}) = 
(0.578304, -0.464099, -8.41439)$ for 
the quasi-one-dimensional periodic boundary case, 
and 
$(\tilde{\alpha},\tilde{\beta},\tilde{\gamma}) = 
(0.534472, -0.449403, -6.76248)$ for 
the quasi-one-dimensional hard-wall boundary case. 
   It may also be noted that the part of the localization widths 
corresponding to the Lyapunov spectrum changing smoothly 
seems to be slightly larger than the value $\mathcal{W}_{ran}$ 
for the localization width for random components 
in the square system with the periodic boundary conditions 
and $d=0.5$ in Fig. \ref{boundary}. 

   In the quasi-one-dimensional periodic boundary case 
in Fig. \ref{boundary} 
we can recognize two pairs of localization widths 
on the line $\mathcal{W}^{(n)}=\mathcal{W}_{wav}$ 
corresponding to the two-point steps of the 
Lyapunov spectra, but there is not a localization width 
of such a value in the hard-wall boundary case. 
   This comes from the fact that in the periodic boundary 
case the system satisfies total momentum conservation 
leading to the two-point steps of the Lyapunov spectrum 
whereas the system with hard-wall boundary conditions 
does not satisfy such a conservation in any direction 
(Detail of this point was discussed in Ref. \cite{Tan02c}.).





\begin{thebibliography}{99}
\bibitem{noteDensi} 
   The density $\rho$ used in this paper is the 
reduced density, and is defined as the ratio of the area 
($N\pi R^{2}$) occupied by hard-disks 
to the total area ($L_{x}L_{y}$). 
   This should be distinguished from another density called 
the particle number density which is defined by 
$\tilde{\rho}\equiv N(2R)^{2}/L_{x}L_{y}$ in our notation. 
   The particle number density has been used in some works 
using many-hard-disk systems with $R=1/2$ \cite{Mil02,For02}. 
   These two kinds of density are simply related by 
$\rho = \pi \tilde{\rho}/4$.
\bibitem{noteWran} 
   There may be exceptional cases in which  
localization widths $\mathcal{W}^{(n)}$ in the Lyapunov 
index corresponding to the continuos part of the Lyapunov spectrum 
is larger than $\mathcal{W}_{ran}$ in very high density cases. 
   See the graph of localization widths for the square system 
in Appendix \ref{AppendixHardwall}.
\bibitem{Gas02note} See Fig. 10 in Ref. \cite{Gas02}. 
   Note that in this figure the Lyapunov exponents are plotted by 
a logarithmic scale, in which the sharp bend of the Lyapunov spectrum 
is represented as a mild step of the Lyapunov spectrum.
\end{thebibliography}
\end{document}